\documentclass[a4paper,11pt]{article}
\usepackage{jcappub} 
\usepackage{lineno}
\usepackage{booktabs}
\usepackage{multirow}

\title{\boldmath Reconstruction of dark energy and late-time  cosmic expansion using the Weighted Function Regression method}

\author{Alex Gonz\'alez-Fuentes}
\author{and Adri\`a G\'omez-Valent}
\affiliation{Departament de F\'isica Qu\`antica i Astrof\'isica, and Institut
de Ciències del Cosmos, \\Universitat de Barcelona, Av.\ Diagonal 647, E-08028 Barcelona,
Catalonia, Spain}

\emailAdd{agonzalezfuentes@icc.ub.edu}
\emailAdd{agomezvalent@icc.ub.edu}

\abstract{Scattered hints of dynamical dark energy (DE) have emerged in various contexts over the past decade. Recent observations from multiple supernova catalogs and the Dark Energy Spectroscopic Instrument (DESI), when combined with CMB data, suggest a highly non-trivial evolution of DE at the $2.5$–$4\sigma$ CL. This evidence is typically quantified using the well-known Chevallier–Polarski–Linder (CPL) parametrization of the DE equation-of-state parameter, $w_{\rm DE}$, which corresponds to a first-order Taylor expansion of $w_{\rm DE}(a)$ around $a = 1$. However, this truncation is to some extent arbitrary and may bias our interpretation of the data, potentially leading us to mistake spurious features of the best-fit CPL model for genuine physical properties of DE. In this work, we apply the Weighted Function Regression (WFR) method to eliminate the subjectivity associated with the choice of truncation order. We assign Bayesian weights to the various orders and compute weighted posterior distributions of the quantities of interest. Using this model-agnostic approach, we reconstruct some of the most relevant cosmological background quantities, namely $w_{\rm DE}(z)$, the DE density $\rho_{\rm DE}(z)$, and several cosmographical functions, including the Hubble function $H(z)$, the deceleration parameter $q(z)$ and the jerk $j(z)$. This allows us to identify which DE features are genuinely preferred by the data, rather than artifacts of a specific parametrization of $w_{\rm DE}(z)$. We examine the robustness of our results against variations in the CMB and SNIa likelihoods. Furthermore, we extend our analysis by allowing for negative DE. Our results corroborate previous indications of dynamical DE reported in the literature, now confirmed for the first time using the WFR method. The combined analysis of CMB, BAO, and SNIa data favors an effective DE component that transitions from phantom to quintessence behavior at redshift $z_{\rm cross}\sim 0.4$. The probability of phantom crossing lies between 96.21\% and 99.97\%, depending on the SNIa data set used, and hence a simple monotonic evolution of the DE density is excluded at the $\sim 2-4\sigma$ CL. Moreover, applying Occam’s razor, we find no significant evidence for a negative dark energy density below $z\sim 2.5-3$. Our reconstructions do not address the Hubble tension, yielding values of $H_0$ consistent with the {\it Planck}/$\Lambda$CDM range. If SH0ES measurements are not affected by systematic biases, the evidence for dynamical dark energy may need to be reassessed.}


\begin{document}
\maketitle
\flushbottom


\section{Introduction}
\label{sec:intro}

The fundamental mechanism behind the current accelerated expansion of the universe \cite{SNIaRiess,SNIaPerlmutter} remains one of the greatest mysteries in physics, holding the key to a deeper understanding of the quantum vacuum and its connection with gravity and the macrocosmos \cite{Zeldovich1968,Weinberg:1988cp,Martin:2012bt,Sola:2013gha,SolaPeracaula:2022hpd}. Although the ultimate nature of the physical entity responsible for this acceleration is still largely unknown, cosmological observations have allowed us to infer some of its basic phenomenological properties \cite{Huterer:2017buf}. There is no doubt that this component violates the strong energy condition $\rho+3p\geq0$ at low redshifts, since its effective\footnote{Effective, in the context of modified gravity theories or coupled dark energy models.} equation of state (EoS) parameter cannot depart excessively from $-1$ and its effective energy density is positive and accounts for roughly 70\% of the universe's current energy budget. Observations also indicate that it does not cluster efficiently -- if at all. 

These properties are trivially realized in the standard model of cosmology, the $\Lambda$CDM model, by a positive cosmological constant $\Lambda$ with associated energy density $\rho_\Lambda=\Lambda/8\pi G\sim  10^{-47}$ GeV$^4$ in natural units, and with negative pressure $p_\Lambda=-\rho_\Lambda$ \cite{Peebles:2002gy,Padmanabhan:2002ji}. Both, $\rho_\Lambda$ and $p_\Lambda$, are rigid, immutable quantities. Despite the theoretical conundrums surrounding $\Lambda$, this model has been very successful in explaining a huge variety of data, ranging from the cosmic microwave background (CMB) anisotropies \cite{WMAP:2012fli,Planck:2018vyg,ACT:2025fju} to the low-redshift data from, e.g., supernovae of Type Ia (SNIa) \cite{SDSS:2014iwm,Pan-STARRS1:2017jku,Scolnic:2021amr}. Nevertheless, it may be only an approximate description of a more complex reality in which the effective energy density and pressure of the component driving the cosmic acceleration evolve with the expansion. Detecting such variation would serve as a clear smoking gun for new physics, as virtually all models of dark energy (DE) and modified gravity introduce dynamics -- either through additional degrees of freedom or novel physical effects within known frameworks (see, e.g., \cite{DEbook} and references therein). Therefore, it is natural for the cosmology community to search for hints of dark energy dynamics in the data, especially in light of the tensions currently affecting the $\Lambda$CDM model, which may already signal a breakdown of the standard cosmological paradigm \cite{Perivolaropoulos:2021jda,DiValentino:2025sru}. 

Indeed, scattered hints of dynamical dark energy have emerged over the years. The first ones appeared (to the best of our knowledge) only five years after the discovery of the accelerated expansion of the universe, using the first available samples of SNIa alone or in combination with the first data release from the Wilkinson Microwave Anisotropy Probe (WMAP)  \cite{Alam:2003fg,Alam:2004jy}. Over the past decade, already in the context of the high-precision cosmology era and amidst the emergence of the Hubble and growth tensions, some groups have reported significant hints of dynamical DE, based on the available extensive background and large-scale structure (LSS) data  \cite{Salvatelli:2014zta,Sola:2015wwa,Sola:2016jky,SolaPeracaula:2016qlq,SolaPeracaula:2017esw,Zhao:2017cud,Sola:2017znb,SolaPeracaula:2018wwm}, see also \cite{Sahni:2014ooa}. The new data from the Dark Energy Spectroscopic Instrument (DESI) \cite{DESI:2024mwx,DESI:DR2constraints} and SNIa samples as Pantheon+ \cite{Scolnic:2021amr}, Union3 \cite{Rubin:2023ovl} and the one from the 5-year data release of the Dark Energy Survey (DES-Y5) \cite{DES:2024jxu,DES:2024hip} have provided further impetus for the investigation of dark energy properties, particularly in characterizing its dynamics. In terms of the famous Chevallier-Polarski-Linder (CPL, a.k.a $w_0w_a$CDM) parametrization of the EoS parameter  \cite{Linder:2002et,Chevallier:2000qy},

\begin{equation}\label{eq:CPL}
w_{\rm DE}(a) = w_0+w_a(1-a)\,,
\end{equation}
which can be thought of as a first-order Taylor expansion of $w_{\rm DE}$ around $a=1$\footnote{Note, however, that it was initially conceived as a fit to minimally coupled scalar field solutions \cite{dePutter:2008wt, Linder:2024rdj}. A large class of models (freezing and thawing) that were spread over the $w\--w'$ plane translated into tight regions in $w_0\--w_a$ space. 
According to this, the use of $w_0\-- w_a$ encodes a physical calibration of the phase space.}, the combination of data on the CMB and baryon acoustic oscillations (BAO) from DESI provides evidence for DE dynamics at a significance level of $3.1\sigma$. Adding the SNIa data from Pantheon+ maintains a high level of evidence, though slightly reduced, at $2.8\sigma$. In contrast, replacing Pantheon+ with DES-Y5 significantly strengthens the signal, raising the significance to $4.2\sigma$ \cite{DESI:DR2constraints}. A similarly strong indication is observed in several data combinations that include BAO data from the Sloan Digital Sky Survey (SDSS) instead of DESI \cite{Park:2024vrw,Gomez-Valent:2024ejh,Giare:2025pzu}. Current observations analyzed under the assumption of the CPL parametrization appear to favor not only a dark energy EoS parameter $w_{\rm DE} \ne -1$ at present, but also a non-trivial evolution of it. Indeed, the preferred CPL model exhibits a crossing of the phantom divide at redshift $z_{\rm cross} \sim 0.3$–$0.5$, transitioning from a phantom regime (at $z > z_{\rm cross}$) to a quintessence-like behavior (at $z < z_{\rm cross}$). See \cite{Malekjani:2024bgi,Giare:2024gpk,Wolf:2025jlc} for recent studies on alternative parameterizations of $w_{\rm DE}(a)$.

One important question is of course whether the physical properties extracted from the data through the lens of the CPL parametrization are directly attributable to the true entity responsible of the cosmic acceleration. It is very important to bear in mind that, in general, once we truncate $w_{\rm DE}(a)$ and constrain its parameters using data that extend beyond the redshift range where the truncated series is valid, the fitting results can no longer be interpreted as accurate estimates of the true values of $w_{\rm DE}(a)$ and its derivatives at the pivot scale factor. This is simply because if the neglected higher-order terms of the expansion are significant, they can induce non-negligible shifts in the fitted parameters. Nevertheless, although the CPL parameterization may admittedly fall short of fully translating  the information contained in the data into a specific microphysical DE model \cite{Shafieloo:2005nd,Wolf:2023uno,Shlivko:2024llw,Wolf:2024eph}\footnote{For instance, one should distinguish between the best fit for $(w_0,w_a)$ obtained within the CPL parametrization and the values  $(w(a=1), -dw/da(a=1))$ of the underlying model. They can be different.}, it might capture some of its relevant phenomenological features \cite{Linder:2002et,dePutter:2008wt,Linder:2024rdj}. It is clear that the evolution of $w_{\rm DE}(a)$ allowed by the CPL makes a crucial difference in the ability of the model to improve the description of the data, since when $w_a$ is set to zero -- and CPL reduces to the well-known $w$CDM parametrization \cite{Turner:1997npq} -- the triad CMB+SNIa+DESI does not provide any compelling hint of DE dynamics \cite{DESI:DR2constraints}. In other words, the CPL is very strongly preferred over the $w$CDM, also when DESI is replaced with SDSS and use is made of LSS data \cite{Gomez-Valent:2024ejh}. However, although higher-order truncations of $w_{\rm DE}(a)$ -- which extend the CPL parametrization by introducing additional parameters -- achieve a similar goodness of fit as the CPL itself, they lead to some notably different conclusions \cite{Nesseris:2025lke, Notari:2024rti}. For example, no statistical evidence is found for a deviation of $w_{\rm DE}$ from $-1$ at the present epoch ($a = 1$) \cite{Nesseris:2025lke}. This highlights the difficulty of disentangling genuine physical features -- those that truly drive a reduction in $\chi^2$ -- from artifacts introduced by subjective choices of models or parametrizations. Clearly, eliminating such subjectivity is crucial for interpreting results reliably and drawing unbiased conclusions that best reflect the underlying physical reality, as informed by current data.

The use of model-independent techniques to reconstruct the cosmological functions of interest can help in this very relevant task. In this paper, we apply the Weighted Function Regression (WFR) method. It was already successfully used in the cosmographical studies \cite{Gomez-Valent:2018hwc,Gomez-Valent:2018gvm}  to reconstruct the Hubble and deceleration parameters making use of low-redshift data, incorporating the information contained in all the relevant truncation orders of the cosmographical expressions. Instead of selecting a single truncation order, the WFR method assigns Bayesian weights -- based on information criteria -- to the contributions of virtually all truncated Taylor expansions. It then constructs the final distribution of the quantities of interest by combining the individual distributions obtained at each truncation order. Thus, the WFR method lets us avoid the subjective choice of a concrete truncation order or function inside a given family by employing Bayesian model-section methods. The WFR method is reminiscent of the Bayesian model averaging approach \cite{Hoeting,Liddle:2006kn, Parkinson:2013hzz}; see also the recent works \cite{Paradiso:2023ohr, Paradiso:2024pcb}.

Herein, we use the WFR method applied in \cite{Gomez-Valent:2018hwc,Gomez-Valent:2018gvm} to reconstruct the effective dark energy EoS and energy density, as well as the deceleration parameter and the jerk, in the light of the CMB data from {\it Planck} (either the 2018 PR3 release or the PR4 release), the BAO data from the second data release of DESI and the Pantheon+ or DES-Y5 SNIa samples. Our method solves the subjectivity issue pointed out in \cite{Nesseris:2025lke} about the use of the CPL parametrization versus those with higher-order truncation terms, and allows us to identify the features that lead to a decrease of the minimum $\chi^2$ with respect to the $\Lambda$CDM and $w$CDM models. Other model-independent methods exist and have been widely used in the literature to reconstruct the DE properties, e.g., Gaussian Processes \cite{Holsclaw:2011wi,Seikel:2012uu,Jesus:2021bxq,Elizalde:2022rss,Escamilla:2023shf,Dinda:2024ktd,Gadbail:2024lek,Yang:2025kgc,DESI:DR2extended,You:2025uon,Mukherjee:2025ytj}, splines \cite{Ye:2024ywg,Ormondroyd:2025exu,Berti:2025phi,Wang:2025vfb,Mukherjee:2025ytj}, the smoothing technique \cite{Shafieloo:2005nd,Shafieloo:2007cs}, neural networks \cite{Mitra:2024ahj}, the binning method alone or combined with principal component analyses \cite{Huterer:2002hy,Crittenden:2005wj,Albrecht:2006mqt,Zhao:2012aw,Liu:2015mkm,Zhao:2017cud,Raveri:2017qvt,Dai2018,Gomez-Valent:2021cbe,DESI:DR2extended,DESI:2025wyn}, symbolic regression \cite{Sousa-Neto:2025gpj} and crossing statistics \cite{DESI:2024aqx,DESI:DR2extended}. Despite being model independent, these methods also require some subjective choices. For example, one has to choose the kernel in Gaussian Processes, the number of bins and their range in the binning method, the order of the Chebyshev expansion in crossing statistics, or the number of knots and polynomial order in the spline interpolation. Different choices can lead to different constraints on the reconstructed functions and therefore can have an impact on the results and conclusions. The Bayesian selection criteria used in the WFR framework can also be applied to these methods, making our approach readily extendable to them as a means of further mitigating the subjectivity problem. This, however, falls beyond the scope of this work.

This paper is organized as follows. In section \ref{sec:WFRmethod}, we explain the WFR method in detail and in section \ref{sec:RecoFunc} we describe the families of expansions that form our WFR-bases. In the first part of the analysis we use a family of Taylor-truncated expansions of the dark energy EoS parameter, while in the second part we directly employ a Taylor-truncated family of the dark energy density. This  allows us to quantify the probability of negative dark energy -- and thus a violation of the two weak energy conditions, $d\rho_{\rm DE}/dz\geq0$ and $\rho_{\rm DE}\geq0$  -- in the past. Such a scenario is very strongly favored by some data sets over the $\Lambda$CDM and is therefore well worth exploring (see, e.g., \cite{Gomez-Valent:2015pia,Malekjani:2023ple,Akarsu:2023mfb,Adil:2023ex,Adil:2023ara,Gomez-Valent:2023uof,Anchordoqui:2024gfa,Gomez-Valent:2024td,Gomez-Valent:2024ejh,Dwivedi:2024okk,Hogas:2025ahb}). In section \ref{sec:data}, we list the individual CMB, BAO and SNIa data sets as well as the various data combinations explored in the paper; section \ref{sec:methodology} is devoted to the description of the methodology. Our results are presented and discussed in section \ref{sec:results} and our conclusions in section \ref{sec:conclusions}. Three appendices complement the information provided in the main text: Appendix \ref{sec:test} is devoted to test the WFR method with mock data; Appendix \ref{sec:contourplots} contains the contour plots obtained in the individual Monte Carlo analyses; and in Appendix \ref{sec:DIC} we compare the impact of different information criteria on the computation of the Bayesian weights and our reconstructions.


\section{The Weighted Function Regression method}\label{sec:WFRmethod}

It has recently been argued that assuming a Taylor expansion of the dark energy equation-of-state parameter and truncating it at first order in $(1-a)$, as in the commonly used CPL parameterization (eq. \ref{eq:CPL}), introduces additional assumptions not supported by the data \cite{Nesseris:2025lke}. As a result, interpreting fitting results based solely on this specific parameterization may lead to biased conclusions. To overcome the problem exposed above, we use the Weighted Function Regression method \cite{Gomez-Valent:2018hwc,Gomez-Valent:2018gvm}. Instead of relying on the results of a particular model with an expansion up to order $J$, the WFR method incorporates the information of all the possible expansions. Let us call $M_0, M_1, ..., M_{n}$ the models linked to the expansions of $w_\mathrm{DE}$ up to order $J=0, 1, ..., n$, in $(1-a)$, respectively\footnote{Other families of functions can be also considered, see section \ref{sec:RecoFunc}; here, however, we base our discussion on expansions of $w_{\rm DE}(a)$ in order to establish a direct connection with the analysis in Ref.~\cite{Nesseris:2025lke}.}. That is, each expansion is a different parameterization of the equation of state of DE, which translates to a different cosmological model. Each model has $n_J=J+1$ parameters describing $w_\mathrm{DE}$.  For $J=0$ one has $w$CDM \cite{Turner:1997npq}, for $J=1$ the CPL \cite{Linder:2002et,Chevallier:2000qy}, and so on. In practice, we only need to consider the expansions up to some order $n$, since, as we will see below, the expansions with higher-order truncations are strongly penalized due to their large number of parameters, which irremediably lead to very small weights. Models with such small weights do not contribute significantly to the weighting process, so they can be safely neglected. 

Let us conceive each expansion as a different possible model and compute the probability density associated to the fact of having a certain shape for a general function $f(z)$\footnote{$f(z)$ can be any cosmological function of interest, or even a concrete parameter. It is important to distinguish this function from the one that characterizes the individual models $M_J$, cf. section \ref{sec:RecoFunc}. These two functions do not necessarily coincide.},
\begin{equation}
P[f(z)|\mathcal{D}]=k\cdot[P(f(z)|M_0)P(M_0|\mathcal{D})+...+P(f(z)|M_{n})P(M_{n}|\mathcal{D})]\,,
\end{equation}
where $\mathcal{D}$ is the data set used in the fitting analysis and $k$ is just a normalization constant that must be fixed by imposing
\begin{equation}
\int[\mathcal{D}f]\,P[f(z)|\mathcal{D}]=1\,.
\end{equation}
Taking into account that 
\begin{equation}
\int[\mathcal{D}f]\,P(f(z)|M_J)=1\quad \forall J\in[0,n]
\end{equation}
and  
\begin{equation}\label{eq:normRel}
\sum_{J=0}^{n} P(M_J|\mathcal{D})=1\,,
\end{equation}
we find $k=1$ and therefore:
\begin{equation}
P[f(z)|\mathcal{D}]=\sum_{J=0}^{n}P(f(z)|M_J)P(M_J|\mathcal{D})\,.
\end{equation}
We now denote $M_*$ as the reference model and rewrite the previous expression as follows,
\begin{equation}\label{eq:penExp}
P[f(z)|\mathcal{D}]=P(M_*|\mathcal{D})\sum_{J=0}^{n}P(f(z)|M_J)\frac{P(M_J|\mathcal{D})}{P(M_*|\mathcal{D})}\,,
\end{equation}
where $\frac{P(M_J|\mathcal{D})}{P(M_*|\mathcal{D})}$ can be identified with the Bayes ratio $B_{J*}$, i.e. the ratio of evidences. The reference model $M_*$ is typically chosen as the one with the largest or the lowest evidence in the whole set $\left\{ M_J\right\}$, although this choice is arbitrary and has no impact on the results, of course. Integrating both sides of eq. (\ref{eq:penExp}) over the measure allows us to find
\begin{equation}
    P(M_*|\mathcal{D})=\left(\sum_{J=0}^{n} B_{J*}\right)^{-1} \,,
\end{equation}
so eq. (\ref{eq:penExp}) can be finally written as
\begin{equation}\label{eq:finExp}
P[f(z)|\mathcal{D}]=\frac{\sum\limits_{J=0}^{n}P(f(z)|M_J)B_{J*}}{\sum\limits_{J=0}^{n}B_{J*}}\,.
\end{equation}
This expression tells us how to reconstruct a function by assigning to each of the models $M_J$ a weight built out from the Bayes factors,

\begin{equation}\label{eq:weight}
W_J = \frac{B_{J*}}{\sum\limits_{I=0}^{n}B_{I*}}\,.
\end{equation}
The effective number of parameters associated to the final reconstruction can be computed simply as follows, 

\begin{equation}\label{eq:Neff}
\mathcal{N}_{\rm eff} = \sum\limits_{J=0}^{n}W_{J}(n_c+n_J)=n_c+\sum\limits_{J=0}^{n}W_{J}n_J\,,
\end{equation}
with $n_c$ the cosmological parameters that are common in all the models of the WFR-basis. 

In practice, the computation of the weights $W_J$ can be strongly simplified by considering the following approximation,

\begin{equation}\label{eq:weightAIC}
    B_{J*} \simeq e^{\Delta {\rm AIC_J}/2}\,,
\end{equation}
where ${\rm AIC}_J = \chi^2_{{\rm min},J} + 2(n_c+n_J)$ is the Akaike information criterion \cite{AIC} and $\Delta {\rm AIC}_J \equiv {\rm AIC}_*-{\rm AIC}_J$. As evident from its definition, the AIC decreases with lower values of $\chi^2_{\rm min}$ and increases with the number of model parameters, thereby penalizing greater model complexity. For a small number of parameters, increasing $n_J$ is generally beneficial, as the resulting decrease in $\chi^2_{\rm min}$ outweighs the penalty term, $2n_J$. However, beyond a certain point, the AIC typically reaches a minimum, making further increases in model complexity unfavorable. In our weighted probability distribution, the contribution of too simple models that do not achieve a small enough reduction of $\chi^2_{\rm min}$ or models that that are excessively complex will be small. The weighted distribution will be dominated by those models around the minimum of AIC. In principle, other information criteria could also be employed, e.g., the deviance \cite{DIC} or the Bayesian \cite{BIC} information criteria. We stick to the use of AIC in our main analyses and compare our results with those obtained with DIC in Appendix \ref{sec:DIC}, where we also explain why BIC does not work properly in this case. AIC and DIC lead to similar weights and, hence, the differences in the reconstructed functions induced by these alternative information criteria are very mild.

In addition to simplifying our analysis from a computational perspective, using these information criteria instead of the exact Bayesian evidences also allows us to avoid the discussion of how prior widths impact the latter \cite{SolaPeracaula:2018wwm,Patel:2024odo}, which lies beyond the scope of this paper. This discussion is inherently challenging (see, e.g., \cite{Cortes:2024lgw,Colgain:2025nzf}), given the limited theoretical guidance available for constructing physically meaningful priors on the parameters that control the shape of our basis functions (cf. section \ref{sec:RecoFunc}). In our analysis, the prior widths are chosen to be sufficiently broad to avoid influencing either the posterior distribution or the inferred values of the information criteria. A more rigorous computation of the WFR weights -- based on exact Bayesian evidences -- could follow the method presented in \cite{Amendola:2024prl}, known as the Frequentist-Bayesian method. This approach uses the distribution of Bayes ratios and is largely insensitive to weak priors. However, it is computationally expensive, especially when considering the full CMB likelihoods, as we do in our study. We therefore opt to leave the exploration of this method in the context of the WFR for future work.  

Before closing this section, we would like to remark that our method allows us to reconstruct, indeed, any function of interest obtainable with an Einstein-Boltzmann solver, not only those that form the WFR function basis. This also includes any function that requires the computation of linear perturbations as the matter power spectrum or the CMB spectra. In this work, though, we focus on the already aforementioned background quantities. We distinguish between basis and reconstructed functions in order to state clearly which are the functions employed to label the models that enter the weighting process, see the next section for further details.


\section{Basis and reconstructed functions}\label{sec:RecoFunc}

The first basis of functions $\{M_J\}$ that we will explore in this work is built by truncating the Taylor expansion of the dark energy EoS parameter around $a=1$ at different orders $J$. Thus, each model $M_J$ of the basis is linked to the following function $w_{{\rm DE},J}(a)$, 

\begin{equation}\label{eq:CPL_gener}
   M_J \Longrightarrow w_{\mathrm{DE},J}(a) = \sum_{l=0}^{J}w_l (1-a)^l\, .
\end{equation}
We denote this basis simply as $w$-basis. In each model $M_J$, $w_{{\rm DE},J}(a)$ can cross $J$ times the phantom divide, at most.  We consider that the effective DE fluid is self-conserved, so the evolution of the DE energy density in model $J$ is ruled by the following conservation equation, 

\begin{equation}\label{eq:consEq}
a\frac{d\rho_{{\rm DE},J}}{da}+3\rho_{{\rm DE},J}[1+w_{{\rm DE},J}(a)]=0\,.
\end{equation}
Using eq. \eqref{eq:CPL_gener} we can integrate the conservation equation analytically and obtain,

\begin{equation}
\rho_{{\rm DE},J}(a) =\rho_{\mathrm{DE},J}^0 f_{{\rm DE},J}(a)\,,
\end{equation}
with $\rho_{\mathrm{DE},J}^0$ the current value of the DE density and

\begin{equation}\label{eq:fDE}
       f_{{\rm DE},J}(a)=a^{-3\left(1+{\sum\limits_{j=0}^{J}w_j}\right)}\exp \left[ -3 \sum_{l=1}^{J} \sum_{k=1}^{l} \binom{l}{k}(-1)^{k}\frac{ w_{l}}{k}(a^{k}-1)\right]\,.
\end{equation}
In the last formula we have used the binomial expansion. The corresponding expression in terms of the redshift is trivially obtained by using the relation $a=(1+z)^{-1}$. The number of extra parameter compared to the standard model is $n_J=J+1$. 

If we keep the terms up to third order in $(1-a)$ in eq. \eqref{eq:fDE}, i.e., if we set $J=3$, and denote $w_1=w_a$, $w_2=w_b$ and $w_3=w_c$ as it is commonly done in the literature, we find,

\begin{equation}
   f_{\mathrm{DE},3}(z)=(1+z)^{3(1+w_0+w_a+w_b+w_c)}\exp \left[ -\frac{3w_az}{1+z}-\frac{3w_b}{2}\frac{(2z+3z^2)}{(1+z)^2}-\frac{w_c}{2}\frac{(6z+15z^2+11z^3)}{(1+z)^3} \right] \, .
\end{equation}
This is the function involved in the computation of the DE energy density of the so-called ${\rm CPL}^{++}$ parametrization \cite{Nesseris:2025lke,Zhang:2017idq,Dai2018}. Obviously, if we set $w_a=w_b=w_c=0$ we retrieve the expression for the $w$CDM; if we do $w_b=w_c=0$ we recover the one for the CPL; and if we just set $w_c=0$ we find the expression for the ${\rm CPL}^{+}$ parametrization \cite{Nesseris:2025lke,Notari:2024rti,Zhang:2017idq,Dai2018,Hussain:2025nqy}. Each of these truncated expansions increases the number of free fitting parameters by one. As will become clear in section~\ref{sec:results}, higher-order truncations ($J \geq 4$) are not necessary for an accurate reconstruction of the functions of interest according to the available data, as their Bayesian weights become small and, hence, do not have a big impact on the results. This is because, beyond a certain point, adding more parameters does not yield a sufficiently significant decrease in $\chi^2_{\rm min}$ to offset the penalization from the model’s increased complexity. This behavior is a general feature in the application of the WFR method: large values of $J$ become effectively irrelevant in the WFR reconstruction \cite{Gomez-Valent:2018hwc,Gomez-Valent:2018gvm}.

From eq. \eqref{eq:consEq} it is evident that $\rho_{\rm DE}$ can become negative only if the EoS parameter diverges at some redshift. This means that our basis functions \eqref{eq:CPL_gener} do not allow us to explore this possibility, which has been shown to be a viable -- and, in some cases, even a preferred--  possibility in light of current data, see, e.g., \cite{Gomez-Valent:2015pia,Malekjani:2023ple,Akarsu:2023mfb,Adil:2023ex,Adil:2023ara,Gomez-Valent:2023uof,Anchordoqui:2024gfa,Gomez-Valent:2024td,DESI:2024aqx,Gomez-Valent:2024ejh,DESI:DR2extended,Dwivedi:2024okk,Hogas:2025ahb}. It is therefore important to analyze a more flexible basis that does not exclude negative values of $\rho_{\rm DE}$ a priori, and to let the data determine whether this is a welcome feature of the effective DE fluid. With this goal in mind, we study a second family of basis functions, that we call $\rho$-basis, obtained by truncating a Taylor series expansion of the DE density around $a = 1$\footnote{For a similar parametrization, but in the pressure, see \cite{Cheng:2025lod}.},

\begin{equation}\label{eq:rhofam}
\widetilde{M}_J\Longrightarrow  \rho_{\mathrm{DE},J}(a) =\rho_{{\rm DE},J}^0 \left[1+\sum_{l=1}^{J}C_l(1-a)^l\right]\, ,
\end{equation}
with the associated EoS parameter, 

\begin{equation}\label{eq:B2w}
    w_{{\rm DE},J}(a) = -1+\frac{a}{3}\left[\frac{{\sum\limits_{l=1}^{J}lC_l(1-a)^{l-1}}}{1+{\sum\limits_{l=1}^{J}C_l(1-a)^{l}}}\right]\,.
\end{equation}
For this second basis, the number of additional parameters with respect to the $\Lambda$CDM is just $n_J=J$. The denominator of the second term in the right-hand side of eq. \eqref{eq:B2w} can have up to $J$ zeros in the range $a\in [0,1]$ depending on the values of the dimensionless $C_l$'s. These zeros generate divergences in $w_{\rm DE}(a)$ and are of course the same zeros of $\rho_{{\rm DE},J}(a)$, cf. eq. \eqref{eq:rhofam}. It is important to remark that the divergences in the EoS do not signal any unphysical behavior in the model, since the DE energy density and pressure, as well all the observable quantities, remain finite $\forall{a}$.

In this work, it will be sufficient to consider the functions \eqref{eq:B2w} up to $J=4$, so we write the expression 

\begin{equation}
w_{{\rm DE},3}(a) = -1+\frac{a}{3}\left[ \frac{C_1+2C_2(1-a)+3C_3(1-a)^2+4C_4(1-a)^3}{1+C_1(1-a)+C_2(1-a)^2+C_3(1-a)^3+C_4(1-a)^4}\right]\,,
\end{equation}
from which it is trivial to obtain the EoS parameters for $n_J=1$, $n_J=2$ and $n_J=3$ by just setting $C_2=C_3=C_4=0$, $C_3=C_4=0$ or $C_4=0$, respectively. 

We note that in the two bases that we will study in this paper, the $w$-basis (eq. \ref{eq:CPL_gener}) and the $\rho$-basis (eq. \ref{eq:rhofam}), the function $w_{\rm DE}(a)$ tends to a constant in the past, i.e., when $a\to 0$. However, there is an important difference. This constant is $-1$ in the $\rho$-basis, while it can differ from this value in the $w$-basis. This means that in the former the DE eventually behaves as a cosmological constant if we go far enough in redshift, whereas in the latter DE can be always dynamical. In practice, though, the behavior of DE at very large redshifts cannot be strongly constrained by the data sets we consider -- namely, CMB, BAO and SNIa, cf. section \ref{sec:data} -- insofar the DE contribution to the total energy and pressure budgets gets diluted fast enough in front of the pressureless matter. We should think of our bases as descriptors of the background DE properties at low redshifts, i.e., at those redshifts at which DE is sensitive to the available data, and do not extrapolate the reconstruction products up to arbitrary remote epochs of the expansion history. This could lead to biased conclusions. We estimate that redshift range in section \ref{sec:results}.

\begin{figure}[t!]
    \centering
    \includegraphics[width=1\linewidth]{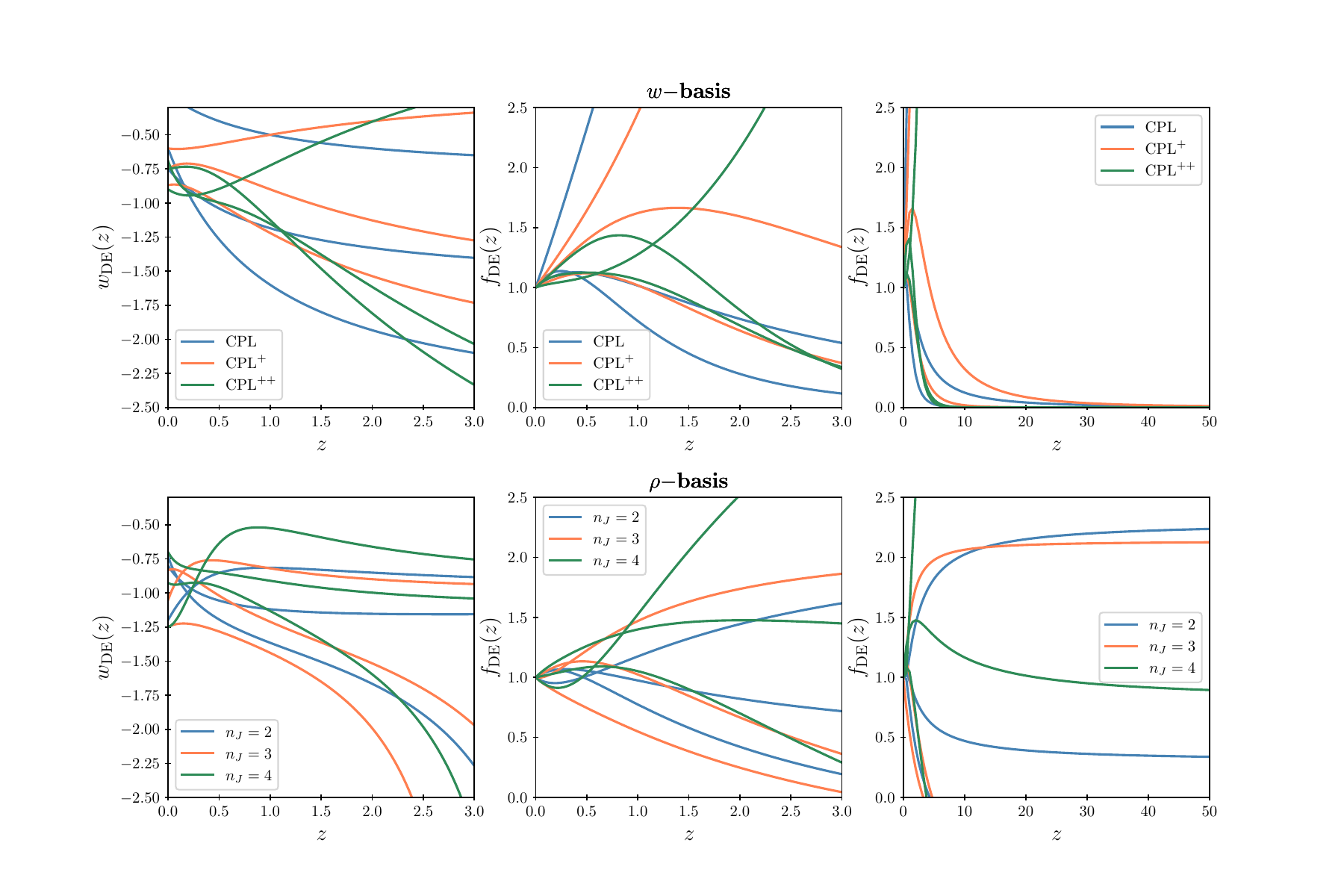}
    \caption{{\it First row:} Representative curves of $w_{\rm DE}(z)$ and $f_{\rm DE}(z)$ for different truncation orders in the $w$-basis. The rightmost plot shows the asymptotic behavior of $f_{\rm DE}$; {\it Second row:} Corresponding curves in the $\rho$-basis. The examples are shown mainly to illustrate the flexibility of the functions in the redshift range of interest, despite their differing asymptotic behavior in the two bases.}
    \label{fig:representative}
\end{figure}

For the sake of visualization, in figure \ref{fig:representative} we display representative forms of $w_{\rm DE}(z)$ and $\rho_{\rm DE}(z)$ obtained from the lowest-degree truncated expansions of the $w$- and $\rho$-bases, which clearly demonstrate that these functions span a wide range of possible phenomenological behaviors of the effective dark energy, providing a flexible framework for describing the underlying physical dynamics in the redshift range of interest.

One could also apply the WFR method using alternative bases, of course, such as, e.g., one built from powers of logarithmic terms, 
$\ln(a)$. However, given the very similar performance of these and other functional forms compared to CPL (cf.~Fig.~4 of \cite{DESI:DR2extended}), 
we decide to restrict the current study to the bases \eqref{eq:CPL_gener} and \eqref{eq:rhofam}, as representative cases. We will see that despite having very different asymptotics, they yield extremely similar reconstructions.

Regarding the DE clustering properties, we assume that it is smooth and cannot form structures, so we set $c_s^2=1$. Therefore,  DE can only influence the LSS processes through its impact on the background. We use the parametrized post-Friedmann approach for the DE to avoid the singularities at the crossing of the phantom divide \cite{Fang:2008sn}.

In addition to the function whose truncations form the basis chosen for the WFR method, we also reconstruct other background cosmological functions of interest. In particular, we reconstruct some cosmographical functions, namely, the Hubble function, 

\begin{equation}\label{eq:E(z)}
    H^2(z) = H_0^2 \left[\Omega^0_{\rm m}(1+z)^3+(1-\Omega^0_{\rm m}) f_{\rm DE}(z)\right] \, ,
\end{equation}
with $\Omega_{\rm m}^0$ the matter density parameter; the deceleration parameter, $q=-\ddot{a}/(aH^2)$ \cite{Sandage1975},

\begin{equation}\label{eq:q(z)}
    q(z)=-1+\frac{1+z}{E(z)}\frac{dE}{dz} \,, 
\end{equation}
with $E(z)=H(z)/H_0$ the normalized Hubble function; and the jerk, $j=\dddot{a}/(aH^3)$ (see, e.g., \cite{Blandford:2004ah}),

\begin{equation}\label{eq:jerk}
    j(z)=q^2(z)+\frac{(1+z)^2}{E(z)} \frac{d^2E}{dz^2} \, .
\end{equation}
These two dimensionless functions are paramount in both theoretical and observational cosmology. They provide explicit information about the late-time kinematics of the universe and can help us understand better the cosmic dynamics. In cosmographical approaches to cosmology their present values, $q_0$ and $j_0$, together with $H_0$, can be used to obtain the cosmological distances as a power series in redshift \cite{Visser:2003vq,Visser:2004bf}. On the other hand, since the universe underwent a transition from deceleration to acceleration, the shape of the second and third derivatives of the scale factor are useful to understand the source behind this evolution, and when the transition happened. Both, $q(z)$ and $j(z)$, are blind to the value of $H_0$ but sensitive to departures from $\Lambda$CDM.  In particular, they depend on the DE density $\Omega_{\rm DE} (z)$ and its EoS $w(z)$. Moreover, the jerk is $j(z)=1$ in the standard model, so a deviation from this value can be interpreted as a hint of new physics. Given the signs of DE dynamics present in the data, it is very interesting to know how the departures from $\Lambda$CDM are imprinted on the kinematics of the universe, and study the problem from a more model-independent perspective. This is why it is useful to show the reconstructed shapes of these functions. 

A brief -- but important -- clarification is in order. In this paper, we reconstruct the universe’s expansion history and the properties of an effective, self-conserved dark energy component -- regardless of its concrete underlying nature. In practice, we treat this component as comprising the energy and pressure content that does not belong to the matter or radiation sectors. The effective dark energy could stem from a modification of gravity, consist of a single or composite fluid, or even be a combination of them (see the discussion in section \ref{sec:results}).


\section{Data Sets}\label{sec:data}

We make use of the following individual data sets:

\begin{itemize}
\item {\it Cosmic Microwave Background}: We use the \textit{Planck} data on the CMB temperature (TT), polarization (EE) and cross (TE) power spectra. In our main fitting analyses, we employ the \texttt{simall}, \texttt{Commander} and \texttt{NPIPE PR4} \texttt{CamSpec} likelihoods \cite{Efstathiou:2019mdh, Rosenberg:2022sdy} for
$\ell<30$ and $\ell\geq 30$, respectively, combined with the likelihood of the \texttt{NPIPE PR4} CMB lensing \cite{Carron:2022eyg}. We will refer to this data set as PlanckPR4, for short. In order to assess the robustness of our reconstructions with respect to the choice of CMB data we use in one of the reconstruction analyses the \texttt{plik}TTTEEE and lensing likelihoods from {\it Planck} PR3 2018 \cite{Planck:2018vyg} instead of the \texttt{NPIPE PR4} high-$\ell$ and lensing likelihoods. We denote the former as Planck18.

\item {\it BAO measurements from DESI}: We use the BAO data from DESI Data Release 2 (DR2), duly accounting for the existing correlations. These data are displayed in Table IV of Ref. \cite{DESI:DR2constraints}.

\item {\it Type Ia supernovae}: It is well-known that DES-Y5 strengthens the statistical evidence for dynamical dark energy compared to Pantheon+ \cite{DESI:2024mwx,DESI:DR2constraints}. Nevertheless, there is an ongoing and heated discussion regarding calibration discrepancies between these two SNIa compilations. Some authors point to the existence of a mismatch in the photometry of nearby $(z \lesssim 0.1)$ SNIa and
those at higher redshift in the DES-Y5 sample \cite{Gialamas:2024lyw,Efstathiou:2024xcq,Notari:2024zmi,Huang:2025som}; see, however, \cite{DES:2025tir}. Given the current status of the discussion and until this issue is not completely settled down, we deem safer to present separate results obtained using the SNIa from DES-Y5 \cite{DES:2024hip,DES:2024jxu} and Pantheon+ \cite{Scolnic:2021amr}. This will allow us to quantify the impact of the SNIa sample choice on the reconstructed functions. 

\end{itemize}
We begin by reconstructing the background cosmological functions described in section \ref{sec:RecoFunc}, using the $w$-basis with both Planck18+DES-Y5+DESI and PlanckPR4+PantheonPlus+DESI. This approach allows us to assess: (i) the impact of upgrading from Planck PR3 to Planck PR4 by comparing the results obtained for the $w$CDM and CPL parameterizations using Planck18+DES-Y5+DESI with those reported by DESI based on PlanckPR4+DES-Y5+DESI \cite{DESI:DR2constraints}; and (ii) the impact of replacing the Pantheon+ SNIa sample with DES-Y5 on both the evidence for dynamical dark energy and the preferred shape of the cosmographical functions extracted from our WFR-reconstruction. Finally, we present the results obtained with the $\rho$-basis using PlanckPR4+PantheonPlus+DESI, which is important to know to what extent negative dark energy densities are allowed at higher redshifts, and also to test the robustness of our reconstruction method under changes in the WFR-basis.


\section{Methodology}\label{sec:methodology}

We use a modified version of the Einstein-Boltzmann code \texttt{CLASS}\footnote{\url{https://github.com/lesgourg/class_public}}\cite{Lesgourgues:2011re,Blas:2011rf} to compute the theoretical predictions of the cosmological observables for the various models studied in this paper, cf. section \ref{sec:RecoFunc}. The exploration of the parameter space through the corresponding Monte Carlo Markov Chain (MCMC) analyses is carried out either with \texttt{MontePython}\footnote{\url{https://github.com/brinckmann/montepython_public}}\cite{Audren:2012w,Brinckmann:2018cvx} or
\texttt{Cobaya}\footnote{\url{https://github.com/CobayaSampler/cobaya}}\cite{Torrado:2020dgo}, depending on the CMB likelihood employed (i.e., {\it Planck} PR3 or PR4 likelihoods, respectively). The convergence of the
chains is tested using the
Gelman-Rubin criterion \cite{GelmanRubin}, with the threshold for chain convergence between $R -1 \leq 0.01$ and $R -1 \leq 0.02$. The chains are analyzed with the \texttt{Python} package \texttt{GetDist}\footnote{\url{https://github.com/cmbant/getdist}} \cite{Lewis:2019xzd} in order to compute the one- and two-dimensional posterior distributions and draw the triangle plots in the planes of interest. This is done after removing the entries of the Markov chains produced in the burn-in phase, which we assume to be $10\%-30\%$ of the total sample. We adopt sufficiently broad flat priors for all the parameters including those entering the basis functions to ensure that the posteriors are essentially shaped by the likelihood. 

Our method basically consists of the following three steps: (i) running of the Monte Carlo analysis with the individual models; (ii) processing of the resulting chains to obtain the constraints of the various parameters and compute the Bayesian weights associated to each model; (iii) reconstructing the functions of interest with the WFR method and computing the weighted constraints of the individual parameters using the formalism detailed in section \ref{sec:WFRmethod}. We have written a \texttt{Python} code that lets us obtain the WFR results from the chains of the individual models\footnote{It will be shared upon reasonable request.}, after removing the burn-in phase. It builds the final chain using the weighted distribution of eq. \eqref{eq:finExp}. Each entry of this chain is obtained by first selecting a model according to the weight \eqref{eq:weight} and then picking randomly one point of the chains associated to that model. Every time we make this step, we store the values of all the parameters at that point of parameter space and use them to compute the cosmological functions of interest on a sufficiently large number of redshift knots. We save the outcome in a chain that we use later on to build the histograms of the reconstructed functions from which we can infer their reconstructed mean values, peaks, and credible intervals.

We consider in all cases a minimal set-up with one massive neutrino of $0.06$ eV and two massless neutrinos. We assume flat spatial hypersurfaces of a Friedmann-Lemaître-Robertson-Walker universe at the background level. Perturbations are considered on top of it. 

We test our pipeline using mock data, assuming $\Lambda$CDM or a more complex DE sector, and obtain in all cases excellent results, which demonstrates the good performance of the WFR method. We display the outcome of these tests in Appendix \ref{sec:test}.

In the tables presenting our fitting results, in the next section, we also report the statistical exclusion level of either the model with the highest $\chi^2_{\rm min}$ in the family of models contained in the basis, the $\Lambda$CDM or both of them, obtained via a likelihood-ratio test \cite{LRtest}. Under the null-hypothesis of this model being the true model in the universe, we know because of Wilks' theorem \cite{Wilk1938} that the distribution of the difference between minimum $\chi^2$ values obtained in that model ($M_A$) and a nested one ($M_B$), i.e., $\Delta\chi^2_{\rm min}=\chi^2_{{\rm min},A}-\chi^2_{{\rm min },B}$, is a $\chi^2_{\nu}$ distribution with $\nu=n_B-n_A$ degrees of freedom. Thus, we can compute the $p$-value associated to the value of $\Delta\chi^2$ obtained in our fitting analysis and determine the confidence level at which we can reject the null hypothesis. As in Ref.  \cite{DESI:DR2constraints}, we translate these $p$-values in number of sigmas, $\xi$,  solving the following equation, 

\begin{equation}\label{eq:pvalue}
p{\rm -value} = 1-\frac{1}{\sqrt{2\pi}}\int_{-\xi}^{\xi}e^{-y^2/2}dy\,.
\end{equation}
For the null hypothesis we will assume the validity of either the $w$CDM model or the $\Lambda$CDM, for which we identify $\xi$ with $E_{w{\rm CDM}}$ or $E_{\Lambda{\rm CDM}}$, respectively.


\section{Results and discussion}\label{sec:results}

In this section, we present and discuss the fitting results obtained with the individual models that form the $w$- and $\rho$-bases. We display them in Tables \ref{tab:tab1}-\ref{tab:tab3}, and figure \ref{fig:indi1}. Then, we use these results to reconstruct the background functions of interest (cf. section \ref{sec:RecoFunc}) with the WFR method using the data sets described in section \ref{sec:data} and the methodology explained in section \ref{sec:methodology}. The main outcome is shown in figure \ref{fig:reco}, and also in Table \ref{tab:tab4}, where we provide the weighted constraints of the main cosmological parameters for the two WFR-bases and data sets explored in this work. For organisational purposes, we discuss the results obtained with Planck18+DES-Y5+DESI in subsection \ref{sec:A1} and those with PlanckPR4+PantheonPlus+DESI in subsection \ref{sec:A2}. 

\begin{table*}
\begin{center}
\begin{tabular}{|c|cccc|}
\multicolumn{5}{c}{\textbf{Planck18+DES-Y5+DESI, $w$-basis}}\\ \hline
Parameter & $w$CDM & CPL & $\mathrm{CPL}^+$ & $\mathrm{CPL}^{++}$ \\ \hline
$10^2\omega_\mathrm{b}$ & $2.255\pm 0.013$ & $2.241\pm 0.013$ & $2.240\pm 0.013$ & $2.239\pm 0.014$ \\ 
$10\,\omega_\mathrm{cdm}$ & $1.117\pm 0.008$ & $1.194\pm 0.008$  & $1.197\pm 0.009$ & $1.196^{+0.010}_{-0.009}$  \\ 
$\ln(10^{10}A_s$) & $3.060^{+0.014}_{-0.017}$ & $3.047\pm 0.015$ & $3.045\pm 0.014$ & $3.044\pm 0.014$ \\ 
$\tau$ & $0.064^{+0.007}_{-0.008}$ & $0.056\pm 0.007$ & $0.055\pm 0.007$  & $0.055\pm 0.007$ \\ 
$n_s$ & $0.972\pm 0.004$ & $0.966\pm  0.004$ & $0.966\pm 0.004$ & $0.966\pm 0.004$ \\ 
$H_0$ [km/s/Mpc] & $67.48\pm 0.51$ & $66.87^{+0.57}_{-0.51}$ &  $67.10\pm 0.60$ & $65.80^{+2.00}_{-0.44}$ \\  
$w_0$ & $-0.964\pm 0.020$ & $-0.748^{+0.042}_{-0.059}$ & $-0.87\pm 0.12$ & $-0.69\pm 0.20$  \\ 
$w_a$ & $-$ & $-0.87^{+0.22}_{-0.17}$ & $0.2\pm 1.0$ & $-2.3^{+2.5}_{-2.9}$ \\ 
$w_b$ & $-$ & $-$ & $-1.8^{+1.9}_{-1.4}$ & $6.9^{+9.5}_{-8.1}$ \\ 
$w_c$ & $-$ & $-$ & $-$ & $-8.3^{+7.4}_{-8.9}$\\ \hline
$\Omega_\mathrm{m}^0$ & $0.309\pm 0.005$ & $0.319^{+0.005}_{-0.006}$ & $0.317\pm 0.006$ & $0.319\pm 0.006$ \\ 
$M$ [mag] & $-19.398\pm 0.010$  & $-19.380\pm 0.011$ & $-19.382\pm 0.012$ & $-19.381\pm 0.011$ \\ 
$\sigma_{12}$ & $0.792\pm 0.007$ & $0.806\pm 0.007$ & $0.807\pm 0.007$ & $0.806^{+0.008}_{-0.007}$\\
$S_8$ & $0.810\pm 0.009$ & $0.832\pm 0.009$ & $0.833\pm 0.009$ & $0.824^{+0.020}_{-0.009}$ \\
$F_{\rm cross}$ [\%] & $0$ & $100.00$ & $99.99$ & $99.99$\\ \hline
$\chi^2_\mathrm{min}$ & 4444.38 & 
4426.60 &  4426.60 & 4426.20  \\ 
$E_{\Lambda{\rm CDM}}$ & $1.70\sigma$ & $4.15\sigma$  & $3.84\sigma$ & $3.61\sigma$   \\
$E_{w{\rm CDM}}$ & $-$ & 4.21$\sigma$ &  $3.81\sigma$ & $3.54\sigma$  \\ 
$\Delta \mathrm{AIC}_J$ & $-$ & $15.78$ & $13.78$  & $12.18$ \\ 
$W_J$ [\%] & $0.02$ & $65.21$ & $23.99$ & $10.78$ \\ \hline
\end{tabular}
\caption{Mean values and uncertainties at 68\% CL obtained using the Planck18+DES-Y5+DESI data set with the individual truncations of the EoS parameter from the $w$-basis (see eq. \ref{eq:CPL_gener}). $F_{\rm cross}$ is the probability of having a crossing of the phantom divide, transitioning from phantom behavior at high redshifts to quintessence at small redshifts, as defined in section \ref{sec:results}. In the last five rows, we report the values of the minimum $\chi^2$, $\chi^2_{{\rm min}}$, the exclusion level of $\Lambda$CDM and $w$CDM in number of sigmas from eq. \ref{eq:pvalue}, the difference of AIC between the model $J$ and the $w$CDM, i.e., $\Delta {\rm AIC}_J={\rm AIC}_{w{\rm CDM}}-{\rm AIC}_{J}$, and the statistical weight used in the WFR, eqs. \eqref{eq:weight}-\eqref{eq:weightAIC}, rounded to two significant figures. The weighted constraints on the various parameters are presented in Table \ref{tab:tab4}. For the constraints on $w_{\rm DE}(z)$ and other background functions, we refer the reader to figures \ref{fig:indi1} and \ref{fig:reco}. The fitting results for the $\Lambda$CDM are displayed in the first column of Table \ref{tab:tabLCDM}.}\label{tab:tab1}
\end{center}
\end{table*}

\subsection{Analysis with Planck18+DES-Y5+DESI}\label{sec:A1}

Let us start analyzing the results obtained with the Planck18+DES-Y5+DESI data set and the various models of the $w$-basis. The constraints for the $w$CDM and CPL models remain close to those obtained with Planck PR4, both in terms of central values and uncertainties, which can be considered a sign of robustness and stability. This can be seen by comparing the fitting results displayed for these models in our Table \ref{tab:tab1} with those reported in Table V of Ref. \cite{DESI:DR2constraints}. For instance, the CPL parameters, $w_0$ and $w_a$, experience only minor shifts of $0.05\sigma$ and $0.03\sigma$, respectively. The Hubble constant is shifted by $0.17\sigma$, while no appreciable shift is observed for $\Omega_{\rm m}^0$. 

\begin{figure}[t!]
    \centering
    \includegraphics[width=0.8\linewidth]{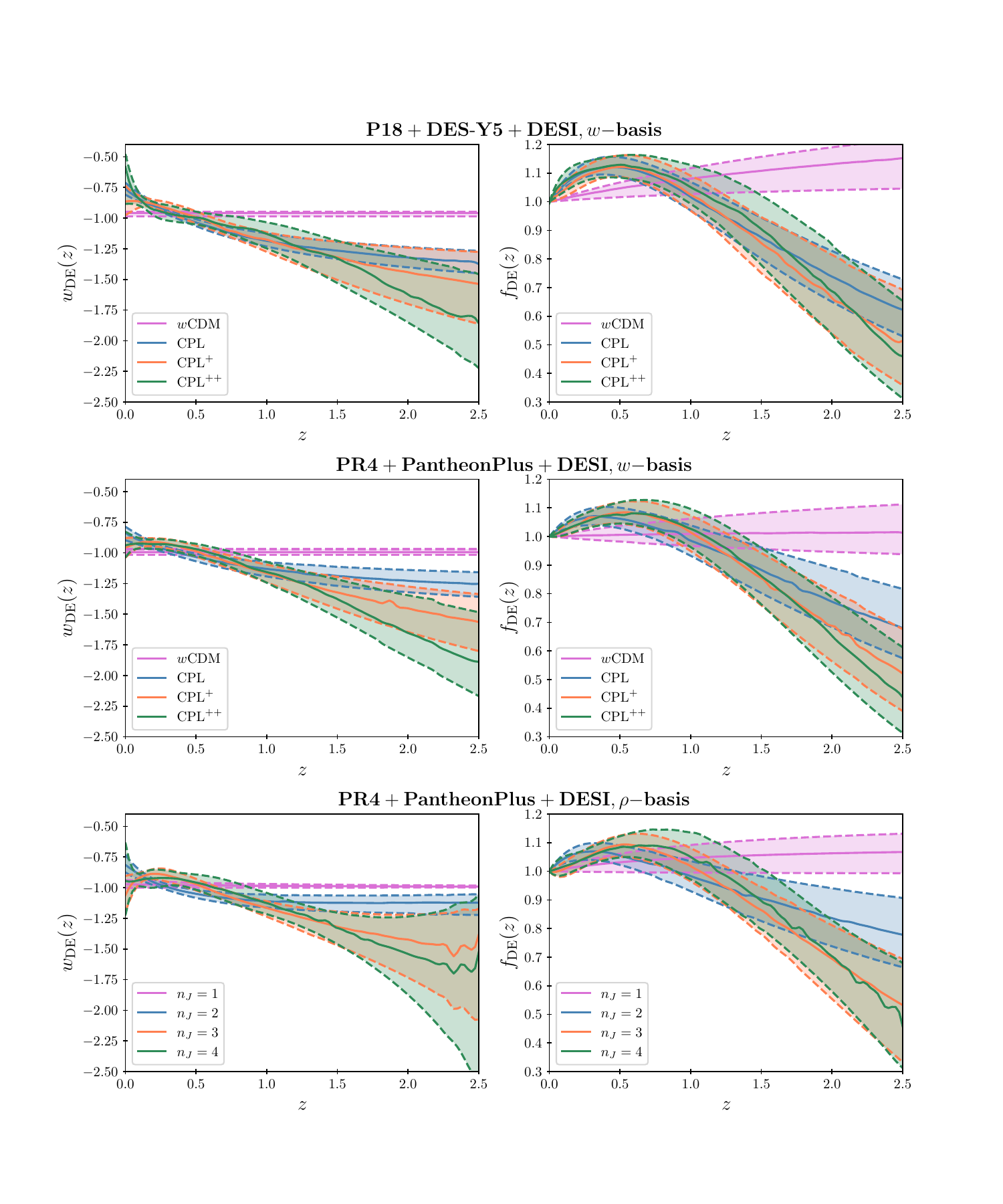}
    \caption{{\it First row:} Reconstructed DE equation of state $w_{\rm DE}(z)$ (left plot) and DE density normalized to its current value $f_{\rm DE}(z)$ (right plot) obtained with Planck18+DES-Y5+DESI for each individual model in the $w$-basis. Solid lines indicate the most probable values, and the shaded areas represent the 68\% confidence intervals; {\it Second row:} The same, but for the analysis with PR4+PantheonPlus+DESI; {\it Third row:} Results obtained with PR4+PantheonPlus+DESI for the various models of the $\rho$-basis.}
    \label{fig:indi1}
\end{figure}

We find a preference for an ever-lasting quintessence-like evolution of the dark energy in the context of the $w$CDM parametrization at $\sim 1.7\sigma$ CL, similar to what was found seven years ago in \cite{SolaPeracaula:2018wwm}. However, this triad of cosmological data strongly favor now a much more complex DE dynamics, as it is evident from the sharp decrease in $\chi^2_{\rm min}$ of almost 18 units found with the CPL with respect to the $w$CDM, which translates into a decrease in AIC of about 16 units. This conclusion is also supported by the likelihood-ratio test, which tells us that the $w$CDM can be rejected at $4.21\sigma$ CL when it is compared to the CPL, and at $3.81\sigma$ CL and $3.54\sigma$ CL when the comparison is performed with the ${\rm CPL}^{+}$ and ${\rm CPL}^{++}$, respectively. The decrease in confidence level found for these models is due to their extra degrees of freedom, which do not manage to ameliorate sufficiently the fit to the data compared to the CPL. 

It is important to understand what concrete features of the DE lead to such a substantial improvement in the description of the cosmological data. To address this question we report the probability of phantom crossing, $F_{\rm cross}$, for each individual model in all our fitting tables (Tables \ref{tab:tab1}, \ref{tab:tab2} and \ref{tab:tab3}) and the weighted reconstruction (in Table \ref{tab:tab4}). We define this phantom crossing as a transition from phantom behavior ($w_{\rm DE}<-1)$ at high redshift to quintessence behavior ($w_{\rm DE}>-1$) in the last stages of the cosmic expansion, i.e., below a given redshift. In practice, $F_{\rm cross}$ can be estimated for each model within a given basis by computing the percentage of realizations in the MCMC chains that lead to the aforesaid crossing of the phantom divide. In the CPL model, we find that 100\% of the realizations undergo this transition. However, when analyzing the CPL alone, we cannot determine whether this result genuinely reflects the preferred dark energy behavior or is instead a consequence of another required feature, compounded by the model's limited flexibility due to its small number of parameters. Therefore, it is useful to also compute the value of $F_{\rm cross}$ in higher-order truncated expansions of the $w$-basis. We find that in both, the ${\rm CPL}^{+}$ and ${\rm CPL}^{++}$ parametrizations, $F_{\rm cross}=99.99\%$, which indicates that the crossing of the phantom divide is a real feature of the DE, according to the data set under study. This is fully aligned with the individual reconstructions of $w_{\rm DE}(z)$ shown in figure \ref{fig:indi1}. Our results for the ${\rm CPL}^{+}$ are consistent with those obtained in \cite{Notari:2024rti} with a very similar data set. See also \cite{Dai2018} for previous results obtained with an older data set.

The constraints on the cosmological parameters common to both the ${\rm CPL}^{+}$ and ${\rm CPL}^{++}$ models are remarkably stable (see again Table \ref{tab:tab1} and the contour plots of figure \ref{fig:contour_rho_panplus1}, in Appendix \ref{sec:contourplots}). The central values and uncertainties of $\omega_{\rm b}$, $\omega_{\rm cdm}$, $\ln(10^{10}A_s)$, $\tau$, $n_s$, $\Omega_{\rm m}^0$, $M$, $\sigma_{12}$ and $S_8$\footnote{In this work, we report constraints on the quantity $\sigma_{12}$, i.e., the current rms mass fluctuations at a fixed scale of $12$ Mpc, instead of $\sigma_8$, since it has been demonstrated to be more representative of the actual amplitude of the power spectrum at linear scales and also because it is more independent from the fiducial value of $H_0$ assumed in galaxy clustering analyses \cite{Sanchez:2020vvb,Forconi:2025cwp}. We also show the posterior values of $S_8$, since this is the optimal observable extracted from weak lensing analyses \cite{Hall:2021qjk,Secco:2022kqg}.} are almost identical to those found with the CPL. We notice little changes in the value of the Hubble parameter, but they remain compatible in all cases, and in the region of small $H_0$, still in strong tension with the SH0ES measurements \cite{Riess:2021jrx}. This is entirely expected, as previous studies have already shown that no late-time dynamical dark energy model can resolve -- or even significantly alleviate -- the Hubble tension \cite{Sola:2017znb,Knox:2019rjx,Krishnan:2021dyb,Lee:2022cyh,Keeley:2022ojz,Gomez-Valent:2023uof}, at least when anisotropic BAO data are taken into account \cite{Gomez-Valent:2023uof}. The stability found in the results of CPL, ${\rm CPL}^{+}$ and ${\rm CPL}^{++}$ might be an indication of the very good performance of the CPL, which under the data set we are considering in this analysis is able to capture the most relevant properties of the DE. In fact, we see that there is no significant gain in the fitting performance of the ${\rm CPL}^{+}$ and ${\rm CPL}^{++}$. This is consistent with the analysis of Ref. \cite{Nesseris:2025lke} -- see their Table I. Indeed, the information criteria penalize the extra degrees of freedom in these models, overcoming the decrease in $\chi^2_{\rm min}$ and making these models to be less favorable compared to the CPL, for which we find the minimum AIC. As we will see below, the reconstructions will be dominated essentially by the CPL, which holds more that 65\% of the probability in the WFR, followed by the ${\rm CPL}^{+}$ ($\sim 24\%$) and ${\rm CPL}^{++}$ ($\sim 11\%$).

\begin{figure}[t!]
    \centering
    {\includegraphics[width=0.4\linewidth]{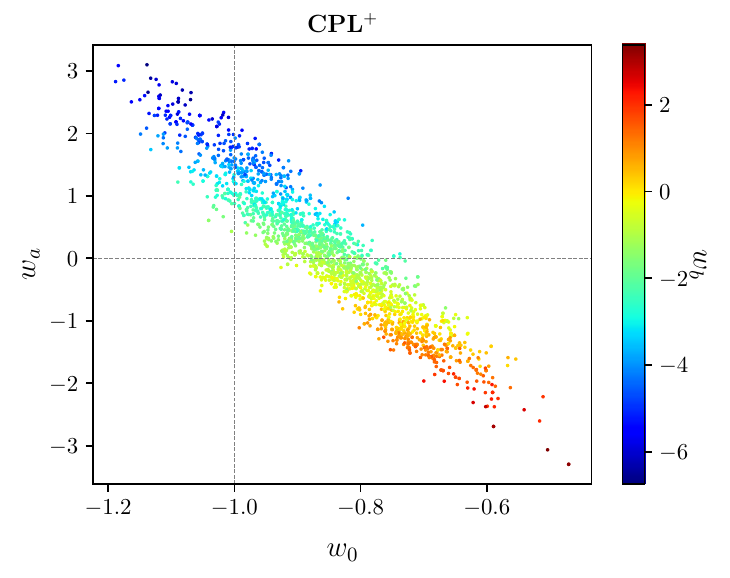}}
    {\includegraphics[width=0.4\linewidth]{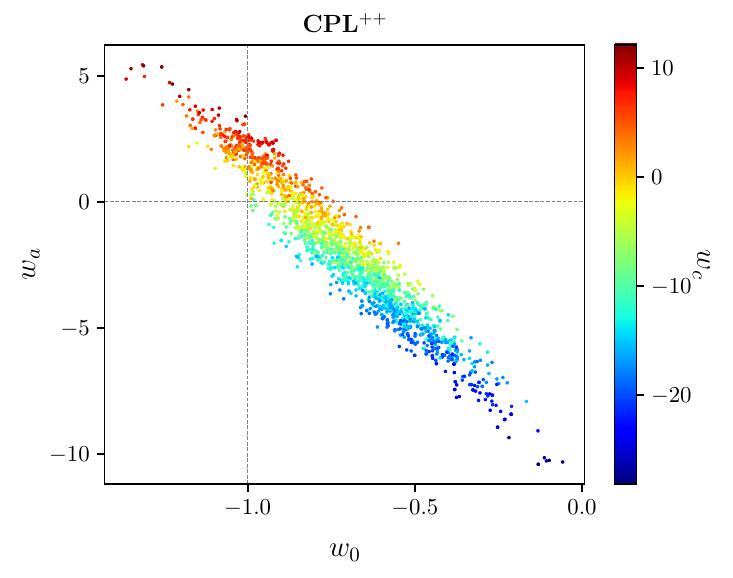}}
    \caption{{\it Left plot:} Scatter plot in the $w_0-w_a$ plane obtained with with Planck18+DES-Y5+DESI for ${\rm CPL}^+$. We also show the corresponding values of the parameter $w_b$ (cf. Table \ref{tab:tab1}). The intersection of grey dashed lines represents the  $\Lambda$CDM point, i.e.,  $(w_0,w_a)=(-1,0)$. {\it Right plot: } Same as in the left plot, but with ${\rm CPL}^{++}$. In this case, the color pallette is for the parameter $w_c$.}
    \label{fig:w0wawb}
\end{figure}

\begin{figure}[t!]
    \centering
    \includegraphics[width=1.02\linewidth]{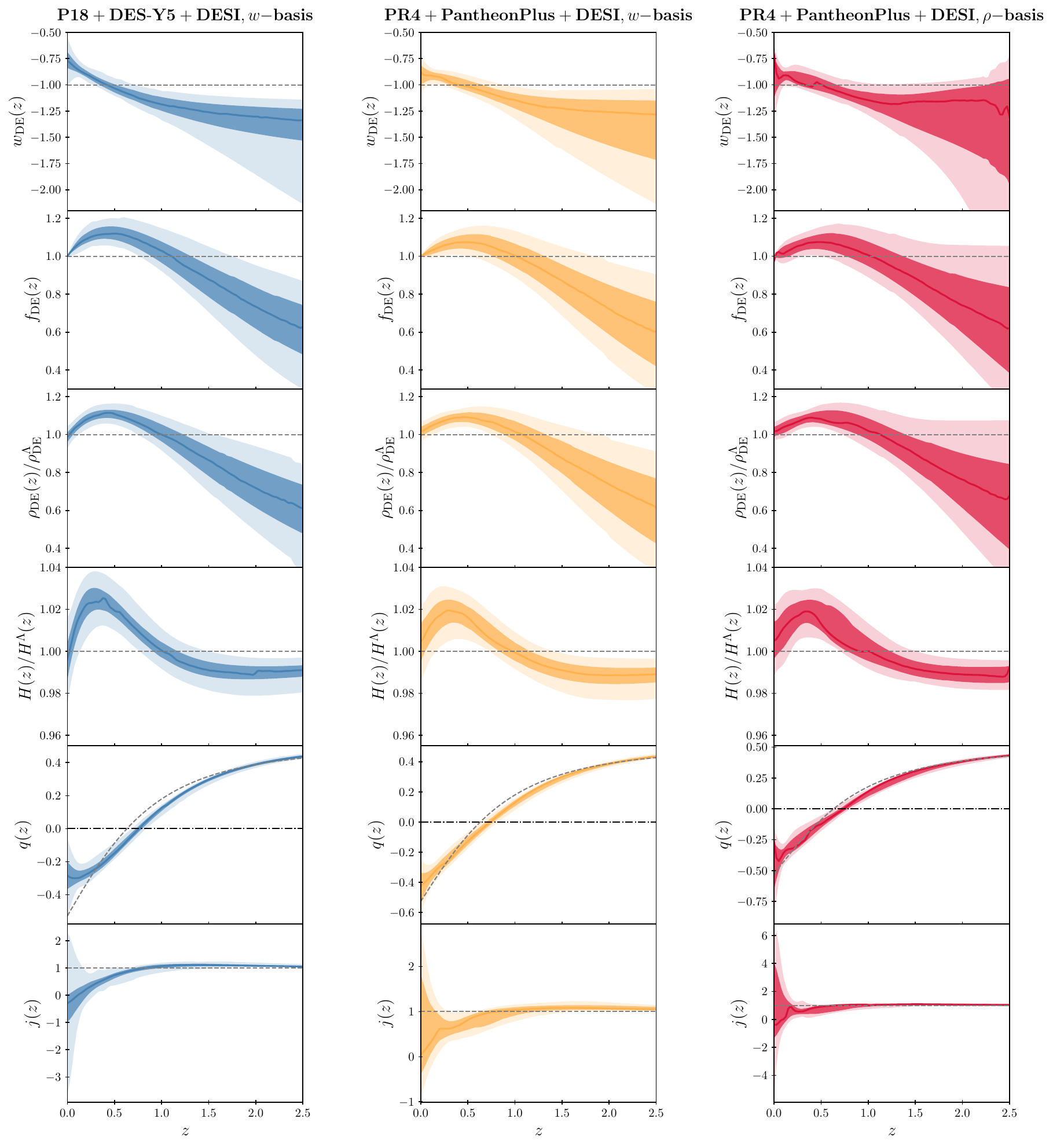}
    \caption{Reconstructed equation of state, energy density normalized to the current value, deceleration parameter and jerk with Planck18+DES-Y5+DESI using the $w$-basis (left-most column), and with PR4+PantheonPlus+DESI using the $w$-basis (central column) and the $\rho$-basis (right-most column). We also plot the reconstructed Hubble rate and energy density normalized to the \textit{Planck} PR4 values for $\Lambda$CDM, for which $\Omega_{\rm m}^0= 0.315$ and  $H_0=67.26 \, {\rm km/s/Mpc}$ \cite{Rosenberg:2022sdy}. The solid colored lines represent the most-probable value and the shaded regions show
    the 68\% and 95\% confidence intervals around it. The grey dashed lines correspond to $\Lambda$CDM values. In the plots of $q(z)$ we also show in black dash-dotted line the border between deceleration ($q>0$) and acceleration ($q<0$) regimes, i.e., $q=0$. Notice that the vertical axis of the last two rows are not aligned for visualization purposes.}
    \label{fig:reco}
\end{figure}

Notice that although the individual -- marginalized one-dimensional -- constraints on the various $w_i$ parameters of the ${\rm CPL}^{+}$ and ${\rm CPL}^{++}$ are compatible with the $\Lambda$CDM values at $1\sigma$ CL, we need to analyze the results in higher dimensional spaces to check if this apparent lack of new physics is real or just a mirage. This is clearly illustrated in the left plot of figure \ref{fig:w0wawb}, where we show that even though the ${\rm CPL}^{+}$ marginalized constraints on $w_0$ ($w_0=-0.87\pm 0.12$) and $w_a$ ($w_a=0.2\pm 1.0$) are compatible with $-1$ and 0, respectively, the contours in the $w_0-w_a$ plane show a clear deviation from the point that defines the standard model $(w_0,w_a)=(-1,0)$. Moreover, we find a strong anti-correlation between $w_a$ and $w_b$, which is somehow expected if the shapes of the reconstructed functions in the ${\rm CPL}^{+}$ have to remain close to those obtained in the CPL, which is actually what figure \ref{fig:indi1} tells us. Something similar happens if we look at the $w_0-w_a$ plane for ${\rm CPL}^{++}$ (right plot of figure \ref{fig:w0wawb}), where the $\Lambda$CDM point now lies at the edge of the 95\% CL band. However, this is achieved at the expense of adding one parameter more to the model and broadening the uncertainties in parameter space. Now there is positive correlation between $w_a$ and the extra parameter $w_c$, with still negative correlation between $w_a$ and $w_b$. In this case, the parameter $w_b$ (not shown in this plot) at the $(w_0,w_a)=(-1,0)$ point is compatible with 0, but this does not mean that the preferred ${\rm CPL}^{++}$ parameterization is consistent with $\Lambda$CDM, since at that point $w_c$ is forced to take values much below 0, close to $-10$. Again, all these correlations and anti-correlations keep the shape of $w_{\rm DE}(z)$ similar to the one found in the CPL.

Nevertheless, we fully agree with the point raised in Ref. \cite{Nesseris:2025lke}. There are features of the CPL parametrization that cannot be taken as representative of the underlying effective dark energy fluid. For instance, imagine we want to answer the following question: What is the evidence of $w_{\rm DE}(z=0)=w_0$ to depart from the $\Lambda$CDM value ($w_0=-1$)? If we evaluate this solely within the CPL framework, the conclusions can be misleading. In the CPL model, the departure of $w_0$ from $-1$ is of about $5\sigma$, suggesting a very significant deviation. However, in the ${\rm CPL}^{+}$ and ${\rm CPL}^{++}$ models we find no statistically significant evidence for such a departure; the results are compatible with $w_0=-1$ at $\sim 1-1.5\sigma$ CL. These differences around $z=0$ are also evident in figure \ref{fig:indi1}. Hence, there are several conclusions that might change a lot depending on the concrete model we choose to parametrize $w_{\rm DE}(a)$.

To address the issue discussed in the last paragraph, we make use of the WFR method. By using the weights displayed in the last row of Table \ref{tab:tab1}, we reconstruct the DE EoS parameter, the DE density and the various cosmographical functions  and present the results in the first column of figure \ref{fig:reco}. See also the constraints on the individual cosmological parameters obtained in our reconstruction, reported in the first column of Table \ref{tab:tab4}. We find only a $\sim 2\sigma$\footnote{Although our constraints are not strictly Gaussian, we express them in terms of the number of sigmas to facilitate comparison using standard terminology.} departure of $w_{\rm DE}(z=0)$ from $-1$. However, we find strong indications in favor of a crossing of the phantom divide, with $F_{\rm cross}=99.97\%$, which corresponds to a signal at $\sim 3.6\sigma$ CL. This value of $F_{\rm cross}$ is slightly smaller than the one found in CPL, ${\rm CPL}^{+}$ and ${\rm CPL}^{++}$ because the $w$CDM has a very small -- but non-null -- weight and in this model $F_{\rm cross}$ is obviously equal to zero. From the reconstructed $w_{\rm DE}(z)$, we find that the redshift at which the crossing of the phantom divide happens is $z_{\rm cross} = 0.40^{+0.10}_{-0.07}$ (most-probable value and 68\% CL uncertainties).

Our WFR-reconstruction has an associated effective number of extra parameters compared to $\Lambda$CDM equal to $\Delta \mathcal{N}_{\rm eff}=2.46$ (cf. eq. \ref{eq:Neff}). When the models are added successively, the associated weights and values of $\Delta \mathcal{N}_{\rm eff}$ change as shown in Table \ref{tab:details}. Assuming that the value of $\chi^2_{\rm min}$ is already almost saturated for $n_J=3$, we can estimate the value of $\Delta \mathcal{N}_{\rm eff}$ we would have for $n_J=4$. We obtain $\Delta \mathcal{N}_{\rm eff}=2.55$, so no important change at the level of the reconstructions coming from the addition of higher truncation orders is expected, and this motivates to consider only models up to $J=3$. This also keeps our methodology viable from the computational point of view.

\begin{table*}[t!]
\begin{center}
\begin{tabular}{|c|ccccc|}
    \hline
    $n_J$ &  $\Delta \mathcal{N}_{\rm eff}$ & $w{\rm CDM}$ & ${\rm CPL}$ & ${\rm CPL}^+$ & ${\rm CPL}^{++}$ \\ \hline
    2& 2.00& $0.04$ & $99.96$ & $-$ & $-$ \\
    3& 2.27 & $0.02$ & $73.08$ & $26.89$ & $-$  \\ 
    4 & 2.46 & $0.02$ & $65.21$ & $23.99$ & $10.78$ \\ \hline
\end{tabular}
\caption{Evolution of the individual model weights and of $\Delta \mathcal{N}_{\rm eff} \equiv \mathcal{N}_{\rm eff} - n_c$ (eq. \ref{eq:Neff}) as models up to $J = 2, 3, 4$ are included in the WFR reconstruction with the $w$-basis, using the Planck18+DES-Y5+DESI data set as an illustrative example. See the main text for further discussion.}\label{tab:details}
\end{center}
\end{table*}

An important aspect to determine is the redshift range over which our reconstruction is sensitive. This might extend beyond the range covered by the low-$z$ data\footnote{The data point with the highest redshift corresponds to the Ly-$\alpha$ DESI DR2 measurement, at an effective redshift of $z_{\rm eff}=2.330$ \cite{DESI:DR2constraints}.}, since CMB might be still sensitive to the particularities of the dark energy component beyond those redshifts, even at the background level. Addressing this question is necessary to determine up to which redshift our reconstructions can be considered reliable. We look for an answer by studying the sensitivity of the angular diameter distance to the last-scattering surface to variations in the shape of $f_{\rm DE}(z)$. This distance is given by the following well-known expression, 

\begin{equation}
D_{A,*}=\frac{c}{1+z_*}\int_0^{z_*}\frac{d\bar{z}}{H(\bar{z})}\,,
\end{equation}
with $z_*$ the redshift at photon decoupling, which is typically close to $z_*\sim 1090$ \cite{Planck:2018vyg}. The Hubble function entering the integral can be actually thought of as a functional of $f_{\rm DE}(z)$, i.e., $H[z,f_{\rm DE}(z)]$, and also depends on the redshift through other functions, as the matter and relativistic energy densities. We can perturb the above expression as follows, 

\begin{equation}
\tilde{D}_{A,*}(z)= \frac{c}{(1+z_*)}\left[\int_0^z\frac{d\bar{z}}{H[\bar{z},f(\bar{z})]}+\int_{z}^{z_*}\frac{d\bar{z}}{H[\bar{z},f(\bar{z})+1]}\right]\,,
\end{equation}
replacing $f_{\rm DE}(z)\to f_{\rm DE}(z)+1$ from a given redshift $z$. By tracking the impact of this perturbation as a function of $z$ and comparing the difference $D_{A,*}-\tilde{D}_{A,*}(z)$ with the typical uncertainty of $D_{A,*}$ extracted from CMB measurements, we can estimate the redshift from which the DE density does not have any effect on the CMB observables given the current sensitivity of the data. Considering the typical uncertainties on the angle $\theta_*$ and the comoving distance $r_s(z_*)$ reported by the Planck collaboration under the assumption of $\Lambda$CDM \cite{Planck:2018vyg}, we find that the uncertainty on $D_{A,*}$ is roughly of $0.02$ Mpc. If we do the aforementioned exercise using the shape of $f_{\rm DE}(z)$ obtained for instance with the mean values of the CPL displayed in Table \ref{tab:tab1}, we find that the condition $D_{A,*}-\tilde{D}_{A,*}(z)<0.02$ Mpc is fulfilled at $z\gtrsim 3$, which is slightly higher than the maximum BAO redshift. This means that our reconstructions can be only trusted below that approximate redshift\footnote{This conclusion is aligned with that of Ref. \cite{Artola:2025zzb}, where the authors show that current data is not very sensitive to the shape of $w(z)$ at sufficiently high redshift.}. In our figure \ref{fig:reco}, though, we only show the reconstructed functions up to $z=2.5$ to ease the visualization of the various curves.

As shown in figure \ref{fig:reco}, our constraint on the current dark energy density, $\rho_{\rm DE}^0$, is fully consistent with the {\it Planck}/$\Lambda$CDM  best-fit value, $\rho_{\rm DE}^{\Lambda}$. However, the reconstructed DE density exhibits a maximum at $z_{\rm cross}$, which significantly exceeds its present value. To maintain consistency with the observed distance to the last scattering surface -- as measured by the CMB under the assumption of standard pre-recombination physics -- $\rho_{\rm DE}(z)$ must fall below $\rho_{\rm DE}^{\Lambda}$ at intermediate redshifts. The same peak and subsequent decrease below the $\Lambda$CDM curve happens also for $H(z)$. As already mentioned, our reconstruction becomes unreliable at redshifts $z\gtrsim 3$, where the data are insensitive to the detailed shape of the DE density due to its suppression relative to the matter density.

Late-time dynamical dark energy alone cannot alleviate the Hubble tension \cite{Sola:2017znb,Knox:2019rjx,Krishnan:2021dyb,Lee:2022cyh,Keeley:2022ojz,Gomez-Valent:2023uof}, and our reconstruction is, of course, no exception. The possibility of an enhanced Hubble expansion rate in the pre-decoupling era --due, e.g., to an energy injection in the dark sector \cite{Poulin:2018cxd,Poulin:2025nfb}, modified gravity effects \cite{SolaPeracaula:2019zsl} or a modified recombination process \cite{Jedamzik:2020krr} -- remains a possibility that can help mitigate the tension. The truth is, though, that no model in the literature is able to efficiently reach the high value of the Hubble parameter measured by SH0ES \cite{Riess:2021jrx} without generating new tensions in other data sectors, at least in the presence of anisotropic (or 3D) BAO data. Moreover, the introduction of early-time new physics tends to weaken the evidence for dynamical dark energy at low redshifts \cite{Poulin:2024ken,Pang:2025lvh,Chaussidon:2025npr,Mirpoorian:2025rfp}. Therefore, a robust determination of the actual level of support for dark energy dynamics at low redshift critically depends on establishing a reliable measurement of $H_0$, which is still a matter of debate in the literature.

The reconstructed deceleration parameter sits on top of the $\Lambda$CDM curve at $z\gtrsim 1.7$ (cf. figure \ref{fig:reco}), but below that redshift important differences appear. The deceleration-acceleration transition redshift ($q(z_{\rm t})=0$) reads $z_{\rm t}=0.77^{+0.03}_{-0.04}$ (most-probable value and 68\% CL uncertainties). This is fully compatible with the model-independent measurement reported in \cite{Gomez-Valent:2018gvm}, $z_{\rm t}\sim 0.8\pm 0.1$, where the WFR method was used with low-redshift data from a combination of SNIa, cosmic chronometers and BAO, and employing cosmographical bases. We find a less decelerated universe than in the standard model before that transition and a more accelerated universe after the transition and up to $z\sim 0.3$. At smaller redshifts, the cosmic speed-up is less aggressive than in $\Lambda$CDM. The present value of the reconstructed deceleration parameter is $q_0=-0.28_{-0.08}^{+0.09}$, which is $\sim 2.4\sigma$ higher than the value obtained in \cite{Gomez-Valent:2018gvm},  $q_0=-0.60\pm 0.10$. It is therefore evident that the deviations from $\Lambda$CDM are also manifest at the kinematical level. Our reconstruction excludes a currently decelerating universe at more than $3\sigma$ CL.

The jerk approaches the $\Lambda$CDM value $j = 1$ at high redshift. This behavior is expected, as dark energy -- whether a cosmological constant or a dynamical component --plays a negligible role well beyond the transition redshift $z_{\rm t}$, due to its dilution relative to matter. However, at low redshift ($z\lesssim 1$) the most-probable reconstructed curve departs from the value predicted by the standard model, even though the uncertainties are still large. Our reconstruction remains consistent with the value $j=1$ at the 95\% CL. 

It is also important to examine how our preferred DE reconstruction impacts the large-scale structure in the universe, particularly in the context of the growth tension \cite{DiValentino:2020vvd}. We find $S_8=0.831^{+0.008}_{-0.010}$ at 68\% CL (cf. table \ref{tab:tab4}). This relatively high value stems from the reduced amount of dark energy at $z\gtrsim 1$ (see figure \ref{fig:reco}), and it is in mild tension with the weak lensing measurements from the Dark Energy Survey (DES) Y3, $S_8=0.775^{+0.026}_{-0.024}$ \cite{DES:2021wwk}, and the Hyper Suprime-Cam (HSC) Y3, $S_8=0.763^{+0.040}_{-0.036}$ \cite{Miyatake:2023njf} -- at the $\sim 2\sigma$ and $\sim 1.7\sigma$ CL, respectively. However, our results remains compatible at $\lesssim 1\sigma$ with the latest data release from the Kilo-Degree Survey (KiDS), $S_8=0.815^{+0.016}_{-0.021}$ \cite{Wright:2025xka}.

A natural question is whether it is possible to construct physical models that can evolve as described above. In the context of minimally coupled single scalar field models, this is not feasible. First, explaining the increase in effective energy density with the cosmic expansion at $z\gtrsim 0.5$ requires the scalar field to be phantom, which in turn necessitates a non-canonical kinetic term, with a sign opposite to the standard one. Notice that, in particular, our reconstruction under this concrete data set (Planck18+DES-Y5+DESI) strongly excludes thawing quintessence models \cite{Caldwell:2005tm}, since at large redshifts $w_{\rm DE}(z)$ can take values much lower than $-1$, and, therefore, DE does not behave as a cosmological constant. This is consistent with the results of \cite{Wolf:2024eph}, although in this analysis we have employed the SNIa from DES-Y5, instead of those from Pantheon+ -- our results obtained with Pantheon+ are provided in section \ref{sec:A2}. Second, these models cannot realize a crossing of the phantom divide. Third, even if such a crossing were possible, it would typically lead to instabilities at the perturbative level.

\begin{figure}[t!]
    \centering
    \includegraphics[width=0.5\linewidth]{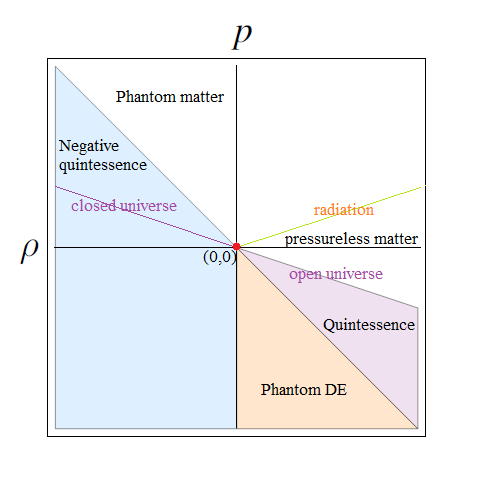}
    \caption{EoS diagram. The pink and orange areas are the standard quintessence ($-1\leq w\leq-1/3$) and phantom ($w\leq-1$) DE regions, respectively. In the white area the strong energy condition is fulfilled, i.e., $\rho+3p\geq 0$ and $\rho+p\geq 0$. In particular, non-relativistic matter ($w=0$) and radiation ($w=1/3$) live there, satisfying $p,\rho\geq 0$. Phantom matter ($w\leq -1$, with $\rho<0$ and $p>0$) occupies the remaining part of the white region, where the strong energy condition holds \cite{Grande:2006nn,Mavromatos:2021urx,Gomez-Valent:2024td,Gomez-Valent:2024ejh}. The contribution of curvature to the Friedmann equations can be thought of as an effective fluid with $w=-1/3$ and $\rho>0$ (open universe) or $\rho<0$ (closed universe). The blue region is the one for species with negative energy density, with decreasing absolute value. Inside that region, the triangle between the ``closed universe'' line and the border of the ``phantom matter'' area can be of phenomenological interest. We coin that form of energy as {\it negative quintessence}, since it has the same EoS as ordinary quintessence ($-1/3 > w> -1$), but has negative energy. See the main text for further details.}
    \label{fig:EoSdiag}
\end{figure}

\begin{table*}
\begin{center}
\begin{tabular}{|c|c|c|}
\hline
Parameter & Planck18+DES-Y5+DESI & PlanckPR4+PantheonPlus+DESI\\ \hline
$10^2\omega_\mathrm{b}$ & $2.251\pm 0.012$ & $2.231^{+0.010}_{-0.009}$  \\ 
$10\,\omega_\mathrm{cdm}$ & $1.181\pm 0.006$ & $1.178\pm 0.006$ \\ 
$\ln(10^{10}A_s$) & $3.056^{+0.013}_{-0.016}$ & $3.047^{+0.013}_{-0.014}$   \\ 
$\tau$ & $0.061^{+0.006}_{-0.008}$ & $0.059^{+0.009}_{-0.008}$  \\ 
$n_s$ & $0.970\pm 0.003$ & $0.968^{+0.003}_{-0.004}$  \\ 
$H_0$ [km/s/Mpc] & $68.35\pm 0.29$ & $68.12\pm 0.27$ \\ \hline
$\Omega_\mathrm{m}^0$ & $0.303\pm 0.004$ & $0.303^{+0.004}_{-0.003}$ \\ 
$M$ [mag] & $-19.387\pm 0.008$ & $-19.420^{+0.009}_{-0.008}$ \\ 
$\sigma_{12}$ & $0.797\pm 0.007$ & $0.794^{+0.005}_{-0.006}$ \\ 
$S_8$ & $0.814\pm 0.008$ & $0.810\pm 0.008$ \\\hline
$\chi^2_\mathrm{min}$ & $4447.26$ & $12394.02$ \\  \hline
\end{tabular}
\caption{Fitting results for the $\Lambda$CDM model obtained using Planck18+DES-Y5+DESI and PlanckPR4+PantheonPlus+DESI. Although we do not include the $\Lambda$CDM in our WFR-bases, we display in Tables \ref{tab:tab1}, \ref{tab:tab2} and \ref{tab:tab3} the statistical tension between the various models considered in section \ref{sec:A2} and the standard model using the likelihood-ratio test (cf. section \ref{sec:methodology} for details). Moreover, it is interesting to see what are the constraints in $\Lambda$CDM, as a reference.}\label{tab:tabLCDM}
\end{center}
\end{table*}

In a two-fluid setup, we can avoid the crossing of the phantom divide. However, simply considering the sum of standard quintessence- and phantom-like fluids with constant EoS parameters -- with positive energy density and negative pressure --  is not sufficient, as they can only produce transitions in the effective (total) dark energy fluid from quintessence to phantom, but not the other way around. This problem can be solved if we consider normal quintessence and a second fluid with negative energy density, with its absolute value decaying with the cosmic expansion sufficiently fast to fall below the quintessence dark energy density at $z\sim 0.5$. This decay has to be slower than the one of non-relativistic matter in order to not spoil the correct expansion history of the universe at large redshifts. Therefore, this fluid should live somewhere in the blue region of the EoS diagram of figure \ref{fig:EoSdiag}, with negative energy density and positive pressure and an equation of state parameter in the range $-1<w<0$. Interestingly, in a closed universe, additional terms in the Friedmann equations effectively behave like those produced by a fluid with $w=-1/3$, satisfying the aforementioned conditions -- apart from some geometric prefactors in the expressions for cosmic distances. Unfortunately, though, curvature is very constrained by CMB, BAO and SNIa data, even when considered alongside dynamical dark energy. The typical upper bound reads, 
$|\Omega_k^0|\lesssim 10^{-3}$ at 68\% CL \cite{DESI:2024mwx}, see also \cite{deCruzPerez:2024shj,Chen:2025mlf}. Hence, quintessence evolving in a closed universe cannot explain our reconstructed results.  However, we cannot discard that models with a combination of regular and negative quintessence fluids -- both with $-1<w<-1/3$, but the latter with negative energy density and positive pressure -- are able to explain the shape of the effective DE evolution observed in the leftmost plots of figure \ref{fig:reco}. Negative quintessence could be described by a minimally coupled scalar field with both negative potential and flipped sign for the kinetic term \cite{NQ}. For a single-field non-minimally coupled scenario, see  \cite{Ye:2024ywg,Ye:2024zpk,Wolf:2025jed}. Other frameworks, like those involving an interaction between dark energy and dark matter \cite{Chakraborty:2025syu} or modified gravity \cite{Odintsov:2024woi,Nojiri:2025low,Mishra:2025goj} have also proved its ability to explain the crossing of the phantom divide. If the combination of Planck18+DES-Y5+DESI is not affected by any significant bias, the measured shape of $f_{\rm DE}(z)$ most probably requires new and exotic physics.

\begin{table*}
\begin{center}
\begin{tabular}{|c|cccc|}
\multicolumn{5}{c}{\textbf{PlanckPR4+PantheonPlus+DESI, $w$-basis}}\\ \hline
Parameter & $w$CDM & CPL & $\mathrm{CPL}^+$ & $\mathrm{CPL}^{++}$  \\ \hline
$10^2\omega_\mathrm{b}$ & $2.232\pm 0.013$ & $2.223^{+0.013}_{-0.014}$ & $2.220\pm 0.013$ & $2.220\pm 0.013$ \\ 
$10\,\omega_\mathrm{cdm}$ & $1.176^{+0.008}_{-0.007}$ & $1.189\pm 0.008$  & $1.192^{+0.008}_{-0.009}$ & $1.192\pm 0.008$ \\ 
$\ln(10^{10}A_s$) & $3.048^{+0.013}_{-0.016}$ & $3.038^{+0.013}_{-0.014}$ & $3.034\pm 0.014$ & $3.034\pm 0.014$ \\ 
$\tau$ & $0.059\pm 0.007$ & $0.053^{+0.007}_{-0.006}$ & $0.051^{+0.007}_{-0.006}$ & $0.052^{+0.006}_{-0.007}$ \\ 
$n_s$ & $0.968\pm 0.003$ & $0.965\pm 0.004$ & $0.965^{+0.003}_{-0.004}$ & $0.964\pm 0.004$\\ 
$H_0$ [km/s/Mpc] & $67.94^{+0.56}_{-0.57}$ & $67.52^{+0.59}_{-0.60}$ & $67.63^{+0.58}_{-0.63}$  & $67.25^{+0.77}_{-0.62}$\\  
$w_0$ & $-0.993^{+0.022}_{-0.023}$ & $-0.842^{+0.055}_{-0.056}$ & $-0.95^{+0.07}_{-0.09}$ & $-0.90^{+0.08}_{-0.13}$\\ 
$w_a$ & $-$ & $-0.59^{+0.22}_{-0.19}$ &$0.54^{+1.02}_{-0.44}$  & $-0.37^{+1.88}_{-0.71}$ \\ 
$w_b$ & $-$ & $-$ & $-1.99^{+0.97}_{-1.42}$ & $1.93^{+3.27}_{-5.26}$ \\ 
$w_c$ & $-$ & $-$ & $-$ & $-4.41^{+4.77}_{-4.52}$ \\ \hline
$\Omega_\mathrm{m}^0$ & $0.305\pm 0.005$ & $0.311\pm 0.006$ & $0.310^{+0.006}_{-0.005}$ & $0.311^{+0.006}_{-0.005}$ \\ 
$M$ [mag] & $-19.424^{+0.010}_{-0.013}$  & $-19.419^{+0.014}_{-0.015}$ & $-19.422^{+0.015}_{-0.014}$ & $-19.423^{+0.015}_{-0.014}$ \\ 
$\sigma_{12}$ & $0.793\pm0.007$ & $0.801^{+0.006}_{-0.007}$ & $0.802\pm 0.007$  & $0.802^{+0.007}_{-0.006}$ \\ 
$S_8$ & $0.809^{+0.008}_{-0.007}$ & $0.822\pm 0.009$ & $0.824^{+0.008}_{-0.009}$ & $0.822^{+0.010}_{-0.008}$ \\
$F_{\rm cross}$ [\%] & $0$ & $99.79$ & $99.92$ & $99.93$ \\\hline
$\chi^2_\mathrm{min}$ & $12394.00$ & $12385.06$  & $12382.30$  & $12382.00$ \\
$E_{\Lambda{\rm CDM}}$ & $0.14\sigma$ & $2.54\sigma$ & $2.64\sigma$ & $2.38\sigma$ \\
$E_{w{\rm CDM}}$ & $-$ & $2.99\sigma$ & $2.98\sigma$ & $2.68\sigma$ \\
$\Delta \mathrm{AIC}_J$ & $-$& $6.94$ & $7.70$ & $6.00$ \\ 
$W_J$ [\%] & $1.00$ & $32.07$ & $46.89$ & $20.04$\\ \hline
\end{tabular}
\caption{Same as in Table \ref{tab:tab1}, but using the PlanckPR4+PantheonPlus+DESI data set.}\label{tab:tab2}
\end{center}
\end{table*}


\subsection{Analysis with PlanckPR4+PantheonPlus+DESI}\label{sec:A2}

We begin this subsection by analyzing the results obtained using the PlanckPR4+ PantheonPlus+DESI data set and the models comprising the $w$-basis, as presented in Table \ref{tab:tab2}. For reference, we show the results obtained with the $\Lambda$CDM in the second column of Table \ref{tab:tabLCDM}. The most significant differences compared to the results discussed in Section \ref{sec:A1} are primarily due to the replacement of the DES-Y5 SNIa data with those from Pantheon+, since we have already seen that the differences between Planck PR3 2018 and PR4 lead to only minor changes in the fitting results. Moreover, the comparison of our results in Table \ref{tab:tab2} for $w$CDM and CPL with those reported in the last columns of Tables IV and V of Ref. \cite{Asorey:2025hgx} indicates that replacing the PR4 \texttt{CamSpec} likelihoods with the \texttt{HiLLiPoP} and \texttt{LoLLiPoP} likelihoods yields no significant shifts in the parameters of these models. Thus, our results are stable under changes in the CMB likelihoods. The DESI collaboration reconstructed the dark energy profile using CMB data (PlanckPR4), BAO data (DESI), and SNIa observations, employing the Union3 sample for the latter \cite{DESI:DR2extended}. We consider it important to also report results based on Pantheon+, as they correspond to the most conservative (or pessimistic) scenario regarding evidence for dynamical DE. This allows for a more comprehensive assessment of the confidence levels at which signs of DE dynamics emerge, upon combining this information with the one provided in the previous section.

\begin{table*}
\begin{center}
\begin{tabular}{|c|cccc|}
\multicolumn{5}{c}{\textbf{PlanckPR4+PantheonPlus+DESI, $\rho$-basis}}\\ \hline
Parameter & $n_J=1$ & $n_J=2$ & $n_J=3$ & $n_J=4$  \\ \hline
$10^2\omega_\mathrm{b}$  & $2.233^{+0.013}_{-0.012}$ & $2.224^{+0.014}_{-0.013}$ & $2.219\pm 0.014$ & $2.220^{+0.013}_{-0.014}$ \\ 
$10\,\omega_\mathrm{cdm}$  & $1.174\pm 0.007$  & $1.187^{+0.008}_{-0.009}$ & $1.193\pm 0.009$ & $1.193^{+0.010}_{-0.009}$ \\ 
$\ln(10^{10}A_s$)  & $3.049^{+0.015}_{-0.014}$ & $3.039\pm 0.014$ & $3.033\pm 0.013$ & $3.032\pm 0.014$ \\ 
$\tau$  & $0.060\pm 0.007$ & $0.055^{+0.006}_{-0.007}$ & $0.051\pm 0.007$ & $0.051\pm 0.007$ \\ 
$n_s$  & $0.969^{+0.003}_{-0.004}$ & $0.965^{+0.004}_{-0.003}$ & $0.964\pm 0.004$ & $0.964\pm 0.004$ \\ 
$H_0$ [km/s/Mpc] & $67.61\pm 0.60$ & $67.46\pm 0.60$ & $67.71\pm 0.64$ & $67.58^{+0.64}_{-0.63}$ \\  
$C_1$   & $0.093^{+0.094}_{-0.101}$ & $0.59^{+0.21}_{-0.22}$  & $-0.16\pm 0.43$ & $0.22^{+0.81}_{-0.79}$\\ 
$C_2$  & $-$ & $-1.25^{+0.47}_{-0.46}$ &  $3.01^{+2.22}_{-2.07}$ & $-0.58^{+6.60}_{-6.67}$ \\ 
$C_3$  & $-$ & $-$ &  $-5.20^{+2.43}_{-2.68}$ & $4.74^{+17.33}_{-18.38}$ \\ 
$C_4$  & $-$ & $-$ &  $-$ & $-8.04^{+14.13}_{-14.59}$ \\ \hline
$\Omega_\mathrm{m}^0$  & $0.307^{+0.006}_{-0.005}$ & $0.311\pm 0.006$ & $0.310\pm 0.006$ &  $0.311\pm 0.006$ \\ 
$M$ [mag] & $-19.430\pm 0.014$ & $-19.420^{+0.015}_{-0.014}$ & $-19.424^{+0.015}_{-0.016}$ & $-19.424^{+0.015}_{-0.014}$ \\ 
$\sigma_{12}$  &  $0.792\pm 0.006$  & $0.799\pm 0.007$ & $0.803\pm 0.007$ & $0.803^{+0.008}_{-0.007}$ \\ 
$S_8$  & $0.809^{+0.008}_{-0.007}$ & $0.820 \pm 0.009$  & $0.825\pm 0.009$ & $0.826^{+0.010}_{-0.009}$ \\
$F_{\rm cross}$ [\%] & $0.00$& $99.44$ & $99.87$ & $99.99$  \\
 $F_{\rm neg}$ [\%] & $0.00$ & $0.00$ & $3.53$ & $3.18$ \\\hline
$\chi^2_\mathrm{min}$  & $12393.52$& $12387.35$&  $12384.27$ &  $12384.20$\\ 
$E_{\Lambda{\rm CDM}}$  & $0.71\sigma$ & $2.10\sigma$  & $2.31\sigma$ & $2.02\sigma$ \\
$\Delta \mathrm{AIC}_J$  & $-$&  $4.17$ & $5.25$ & $3.32$ \\ 
$W_J$ [\%]  & $3.56$ & $28.62$ & $49.12$& $18.71$ \\  \hline
\end{tabular}
\caption{Same as in Table \ref{tab:tab2}, but using the $\rho$-basis, eq. \eqref{eq:rhofam}. Here, we also display the values of $F_{\rm neg}$, i.e., the probability of the DE density to take negative values in the redshift range $z\in [0,3]$.}\label{tab:tab3}
\end{center}
\end{table*}

As expected, the use of the Pantheon+ sample leads to a decrease of the evidence for dynamical DE. However, substantial hints in favor of it remain. In this case, the simple, ever-lasting monotonic evolution of $\rho_{\rm DE}(z)$ permitted by the $w$CDM model is excluded at approximately $3\sigma$ CL in favor of the CPL -- compared to the $4.21\sigma$ exclusion obtained using DES-Y5 data. The exclusion of $\Lambda$CDM with respect to CPL that results from the likelihood-ratio test is instead of about 2.5$\sigma$. It is a bit lower than the one found using $w$CDM since $\Lambda$CDM performs very similarly as the $w$CDM, but has one parameter less. What happens if we truncate the Taylor expansion at higher orders?   The statistical significance remains stable in CPL${^+}$ compared to CPL, but decreases in CPL$^{++}$ to $2.3\sigma$ and $2.5\sigma$ CL relative to $\Lambda$CDM and $w$CDM, respectively. For a comparison of the individual shapes of the EoS parameter and the DE density at $68\%$ CL, see the plots in the second row of figure \ref{fig:indi1}.

\begin{table*}
\begin{center}
\begin{tabular}{|c|ccc|}
\multicolumn{4}{c}{\textbf{Reconstructed parameters}}\\ \hline
\multirow{2}{*}{Parameter} & \textbf{P18+DES-Y5} & \textbf{PR4+PantheonPlus} & \textbf{PR4+PantheonPlus} \\
  & \textbf{$w$-basis} & \textbf{$w$-basis} & \textbf{$\rho$-basis} \\
\hline
$10^2\omega_\mathrm{b}$ & $2.240^{+0.014}_{-0.013}$ &$2.222\pm 0.013$ & $2.222\pm 0.014$ \\ 
$10\,\omega_\mathrm{cdm}$ & $1.195^{+0.009}_{-0.008}$ &  $1.191\pm 0.009$ & $1.191\pm 0.010$ \\ 
$\ln(10^{10}A_s$) & $3.045^{+0.015}_{-0.014}$ & $3.035\pm 0.014$ & $3.035 \pm 0.014$ \\ 
$\tau$ & $0.056\pm 0.007$ & $0.052\pm 0.007$ & $0.052\pm 0.007$ \\ 
$n_s$ & $0.966^{+0.004}_{-0.003}$ & $0.965\pm 0.004$ & $0.965\pm 0.004$ \\ 
$H_0$ [kms/s/Mpc] & $66.81^{+0.47}_{-0.72}$ & $67.52_{-0.62}^{+0.63}$ & $67.61 \pm 0.63$ \\  \hline
$\Omega_\mathrm{m}^0$ &  $0.318^{+0.006}_{-0.005}$  & $0.311\pm 0.006$ & $0.311\pm 0.006$\\ 
$M$ [mag] & $-19.381\pm 0.011$  & $-19.421\pm 0.015$ & $-19.423\pm 0.015$\\ 
$\sigma_{12}$ & $0.806\pm 0.007$ & $0.801\pm 0.007$ & $0.801\pm 0.008$\\ 
$S_8$ & $0.831^{+0.008}_{-0.010}$ & $0.823\pm 0.009$ & $0.823\pm 0.010$ \\
$z_{\rm t}$ & $0.77^{+0.03}_{-0.04}$ & $0.74\pm 0.04$ & $0.74\pm 0.04$\\
$F_{\rm cross}$ [\%] & $99.97$ & $98.88$ & $96.21$\\
$F_{\rm neg}$ [\%] & $-$ & $-$ & 2.32\\
$z_{\rm cross}$ & $0.40^{+0.10}_{-0.07}$ & $0.38_{-0.08}^{+0.23}$ & $0.57^{+0.09}_{-0.32}$ \\
$\Delta \mathcal{N}_{\rm eff}$ & $2.46$ & $2.86$ & $2.83$\\\hline

\end{tabular}
\caption{Weighted constraints on the various parameters for the different data sets and WFR-bases employed. To save some space, we omit the label DESI, which is common in all three combinations. We also display constraints on the deceleration-acceleration transition redshift, $z_{\rm t}$, the redshift of phantom crossing, $z_{\rm cross}$, the probability of phantom crossing, $F_{\rm cross}$, the probability of having negative DE in the redshift range $z\in [0,3]$, $F_{\rm neg}$, and the effective number of extra parameters, $\Delta \mathcal{N}_{\rm eff}\equiv \mathcal{N}_{\rm eff}-n_c$ (eq. \ref{eq:Neff}), for the weighted reconstructions. We show the mean and 68\% uncertainties for all the quantities, except for $z_t$ and $z_{\rm cross}$ for which we show the mode instead of the mean.}\label{tab:tab4}
\end{center}
\end{table*}

As shown in the analysis of section \ref{sec:A1}, the constraints on the cosmological parameters shared across all models are remarkably stable for CPL, CPL$^+$ and CPL$^{++}$, also with Pantheon+, which again indicates that CPL is already able to capture the most important features of the underlying effective DE present in our data set. It is important to remark, though, that the constraints that one obtains for, e.g., $w_{\rm DE}(z)$ or $f_{\rm DE}(z)$, using only one of these parametrizations  differ from the final reconstructed result (cf. again figures \ref{fig:indi1} and \ref{fig:reco}). The reconstructed uncertainties grow due to the impact of the higher-order parameterizations, which have larger error bars. The WFR method comes to the rescue, getting rid of this subjectivity problem by letting the data choose the most optimal weighted distribution through the use of Bayesian tools. The shapes of all the reconstructed functions displayed in the central plots of figure \ref{fig:reco} are quite similar to those found with DES-Y5. This similarity is obvious at naked eye, but let us be more quantitative. For instance, we find that the transition from a decelerated to an accelerated universe happens at $z_{\rm t}=0.74\pm 0.04$ at 68\% CL, in full accordance with the result found with DES-Y5. The SNIa from Pantheon+ prefer a  more accelerated universe at present, although the change is not statistically significant at all. We still find a peak in $f_{\rm DE}(z)$ at redshift $z\sim 0.5$. However, with Pantheon+ the $\Lambda$CDM value $f^\Lambda=1$ falls within the 95$\%$ CL bands in practically all the redshift range covered by our WFR reconstruction, and the value of $w_{\rm DE}(z=0)$ becomes compatible with $-1$ at only $1\sigma$ CL. The reconstructed probability of crossing of the phantom divide is now $F_{\rm cross}=98.88\%$ -- see the individual values of $F_{\rm cross}$ obtained in each model of the $w$-basis in Table \ref{tab:tab2}. It is slightly smaller than the value found using DES-Y5, due to weaker evidence for dynamical dark energy in that SNIa sample. The constraint on the redshift at which this crossing occurs reads, $z_{\rm cross}=0.38_{-0.08}^{+0.23}$, again in perfect agreement with the value reported in section \ref{sec:A1}.

It is important to test the robustness of the WFR method by considering alternative bases. To this end, we use the $\rho$-basis \eqref{eq:rhofam} and compare the results with those obtained using the $w$-basis \eqref{eq:CPL_gener}, employing the same data set: PlanckPR4+PantheonPlus+DESI. The main reason for choosing the $\rho$-basis is that it has enough flexibility to allow the DE density to become negative at high redshifts. Hence, this basis helps us quantify not only the probability of crossing the phantom divide, $F_{\rm cross}$, but also the probability of having negative DE, denoted as $F_{\rm neg}$. Both features lead to a violation of the weak energy condition.

The individual fitting results for the various models entering the WFR reconstruction are reported in Table \ref{tab:tab3}, and the corresponding curves of the EoS parameter and the DE density are presented in the bottom plots of figure \ref{fig:indi1}. These plots look very similar to those obtained with the $w$-basis, when the same number of additional parameters are employed. The fair comparison should be performed between the $w$CDM, CPL, CPL$^{+}$ and CPL$^{++}$ models of the $w$-basis and the models with $n_J=1,2,3,4$ of the $\rho$-basis, respectively. This similarity is imprinted of course also on the weights for the individual models and the shape of the final reconstructions, see the right-most plots in figure \ref{fig:reco}.

Some of the realizations obtained with $n_J=3$ and $n_J=4$ in our Markov chains lead to divergences of the EoS parameter at $z\gtrsim 2$ due to the transition from positive to negative values of the DE density, although of course all the physical quantities -- including $\rho_{\rm DE}(z)$ -- remain finite. The individual values of $F_{\rm neg}$ are in all cases below 3.6\% and the weighted probability of having $\rho_{\rm DE}(z)<0$ in the redshift range $z\in[0,3]$ is $F_{\rm neg}=2.32\%$, which is small, but still non-zero. Models with a sudden transition at $z\sim 2$ from negative to positive values of $\rho_{\rm DE}$ as the so-called $\Lambda_s$CDM or $w$XCDM \cite{Gomez-Valent:2024ejh,Gomez-Valent:2024td} can describe the data also much better than the $\Lambda$CDM and the $w$CDM, but perform a bit worse than CPL when anisotropic (or 3D) BAO data as those employed in this work are employed in the fitting analysis -- cf. Table 1 of Ref. \cite{Gomez-Valent:2024td}\footnote{In the presence of angular (or 2D) BAO, which are in tension with 3D BAO \cite{Camarena:2019rmj,Gomez-Valent:2023uof,Favale:2024sdq}, the situation is completely different. Models exhibiting a sharp transition in $\rho_{\rm DE}(z)$ as $\Lambda_s$CDM \cite{Akarsu:2023mfb}, bimetric gravity \cite{Dwivedi:2024okk} or, more conspicuously, the $w$XCDM \cite{Gomez-Valent:2024ejh,Gomez-Valent:2024td}, improve by a lot the performance of the CPL parametrization, cf. again Table 1 in Ref. \cite{Gomez-Valent:2024td}.}.

The probability of phantom crossing in the analysis with the $\rho$-basis is of $96.21\%$, with $z_{\rm cross} = 0.57^{+0.09}_{-0.32}$. The deceleration-acceleration transition happens at $z_{\rm t}=0.74\pm 0.04$,  and the jerk parameter is compatible with $j^\Lambda=1$ at $<1\sigma$ CL. All these results are again fully compatible with those obtained with the $w$-basis, which is a clear sign of the robustness of the WFR method. Both bases are able to capture the most important aspects that are required for an appropriate description of the data -- the negative DE feature does not play an important role. In fact, we find with both bases an almost identical value of the effective number of extra parameters $\Delta \mathcal{N}_{\rm eff}\sim 2.8$, slightly larger than in the analysis with DES-Y5, and the differences in the weighted constraints for the cosmological parameters for the $w$- and $\rho$-bases are also derisory -- cf. Table \ref{tab:tab4}. Finally, the replacement of DES-Y5 with Pantheon+ does not change at all the status of the Hubble and growth tensions. Therefore, the related comments made in section \ref{sec:A1} still hold.


\section{Conclusions}\label{sec:conclusions}

One of the greatest open questions in cosmology is the nature of dark energy -- the entity responsible for the accelerated expansion of the universe at late times. Determining whether this entity evolves with the cosmic expansion is a natural step toward answering this fundamental question, and is therefore a worthwhile direction for further investigation. Hints of dynamical dark energy have appeared intermittently in the literature, arising in various contexts and from diverse perspectives. Although the origin and implications of these signals for dark energy dynamics may have shifted with recent observations, it is important to acknowledge the contributions of earlier studies that explored this line of research, see, e.g.,  \cite{Alam:2003fg,Alam:2004jy,Shafieloo:2007cs,Salvatelli:2014zta,Sahni:2014ooa,Sola:2015wwa,Sola:2016jky,SolaPeracaula:2016qlq,SolaPeracaula:2017esw,Zhao:2017cud,Sola:2017znb,SolaPeracaula:2018wwm}. Now, the combination of CMB data from {\it Planck}, BAO data -- either from BOSS/eBOSS or DESI \cite{Park:2024vrw,Gomez-Valent:2024ejh,Giare:2025pzu} -- and different SNIa samples lead to substantial evidence in favor of the dynamical evolution of the effective DE. Quantified in terms of the Chevallier-Polarski-Linder (CPL) parametrization \cite{Linder:2002et,Chevallier:2000qy} -- a first-order Taylor expansion of the equation-of-state parameter around $a=1$ --, this evidence lies between the $\sim 2.5\sigma$ and $\sim 4\sigma$ CL depending on whether the SNIa from Pantheon+ or DES-Y5 are used in the fitting analysis \cite{DESI:DR2constraints}. However, understanding the extent to which this statistical evidence depends on the specific parametrization employed is of utmost importance to ensure that the features extracted from the data are genuine, and not artifacts introduced by the parametrization itself \cite{Nesseris:2025lke}.

In this work, we have employed a largely model-agnostic reconstruction technique known as the Weighted Function Regression (WFR) method, also used in Refs. \cite{Liddle:2006kn,Gomez-Valent:2018hwc,Gomez-Valent:2018gvm}, to mitigate the subjectivity associated with choosing a specific parametrization or truncation order in series expansions. We have applied it to the reconstruction of the effective dark energy background properties, assuming that this component is covariantly conserved and that it does not cluster efficiently. We also reconstruct the most relevant cosmographical functions. Our results are robust under the choice of WFR-basis. They are aligned with those reported by DESI \cite{DESI:DR2extended} and other groups using different methods. The combined analysis of CMB from {\it Planck}, BAO from DESI, and SNIa  data from DES-Y5 or Pantheon+ favors an effective DE component that transitions from phantom to quintessence behavior at redshift $z_{\rm cross}\sim 0.4$. The probability of phantom crossing lies between 96.21\% and 99.97\%, depending on the SNIa data set used, so we can exclude a monotonic evolution of the DE density at $\sim 2-4\sigma$ CL. Recent results in the literature point to similar levels of evidence \cite{Ozulker:2025ehg,Keeley:2025rlg,Silva:2025twg}. In \cite{Keeley:2025rlg}, for instance, the authors conclude that the probability of a detection of a spurious phantom crossing is $3.2\%$, based on the Union3 SNIa sample. This value lies between our estimates using Pantheon+ ($1.1\%$–$3.8\%$). Only when Pantheon+ is replaced with DES-Y5 does
the signal exceed 3$\sigma$. 

Applying Occam’s razor, we do not find any significant hint for a negative dark energy density below $z\sim 2.5-3$. Our reconstructions lead to small values of $H_0$, in the range preferred by {\it Planck} \cite{Planck:2018vyg,Rosenberg:2022sdy} and ACT \cite{ACT:2025fju} under the assumption of $\Lambda$CDM. This result is not unexpected, as no late-time dynamical dark energy model has been shown to resolve the Hubble tension when anisotropic BAO data are included in the fitting analyses \cite{Sola:2017znb,Knox:2019rjx,Krishnan:2021dyb,Lee:2022cyh,Keeley:2022ojz,Gomez-Valent:2023uof}. If the SH0ES measurements are not significantly affected by systematic biases, the current evidence for dynamical dark energy may need to be reconsidered. It will also be important to closely follow the ongoing discussion regarding systematics in the SNIa samples and their influence on the results, as these clearly play a non-negligible role in the analysis.

Explaining the preferred shape of the dark energy density in the simplest case of minimally coupled scalar fields requires the presence of both a standard and a negative quintessence fields \cite{NQ}. Other possibilities include scenarios of modified gravity or interactions in the dark sector, see, e.g. \cite{Ye:2024ywg,Ye:2024zpk,Wolf:2025jed,Chakraborty:2025syu,Odintsov:2024woi,Nojiri:2025low,Mishra:2025goj}.

Future research directions include a more rigorous calculation of the Bayesian weights, following the methodology of \cite{Patel:2024odo}, as well as a reconstruction of the late-time dark energy properties that takes into account potential new physics in the pre-recombination era -- an aspect that may be critical for a more comprehensive understanding of the Hubble tension and its impact on the quantification of the evidence for evolving dark energy.

We are living exciting times for cosmology. Upcoming data releases from DESI, along with future observations from Euclid, may help to consolidate these preliminary hints of dynamical dark energy and, hopefully, bring us a step closer to understanding the fundamental mechanism behind the entity that constitutes approximately 70\% of the universe’s current energy budget. We believe that the model-agnostic tool employed in this paper will contribute to this important task.


\appendix

\section{Testing the WFR method with mock data}\label{sec:test}
We validate our pipeline with mock data for $\Lambda$CDM and an extreme thawing model exhibiting a sudden transition from $w(z)\gtrsim -1$ at $z>0.5$ to $w(z=0)=-0.5$. For the latter, we use the algebraic expression proposed in \cite{Linder:2007wa},

\begin{equation}\label{eq:algebraicthwaing}
    w(a)=-1+(1+w_0)a^p\left(\frac{1+b}{1+ba^{-3}}\right)^{1-p/3} \, ,
\end{equation}
with $b=0.5$, $w_0=-0.5$ and $p=20$, as in Appendix E of \cite{DESI:DR2extended}. For both the $\Lambda$CDM and the thawing quintessence model we fix $\omega_{\rm b}=0.02223$, $\omega_{\rm cdm}=0.1192$ and $100\theta_*=1.04103$. We do not add any noise in the mock data since our aim is to test whether the method itself leads to any bias in the reconstruction process.

For simplicity and to speed up the calculations, we construct the mocks using compressed information from the early universe instead of the full CMB likelihoods in the form of a correlated multivariate Gaussian on $(\theta_*, \omega_{\rm b}, \omega_{\rm bc})_\mathrm{CMB}$ \cite{Lemos:2023xhs}, as done in \cite{DESI:DR2constraints}, in combination with the corresponding mock data for BAO and SNIa using the redshifts and covariance matrices from DESI DR2 and DES-Y5, respectively.

\begin{figure}[t!]
    \centering
    \includegraphics[width=0.9\linewidth]{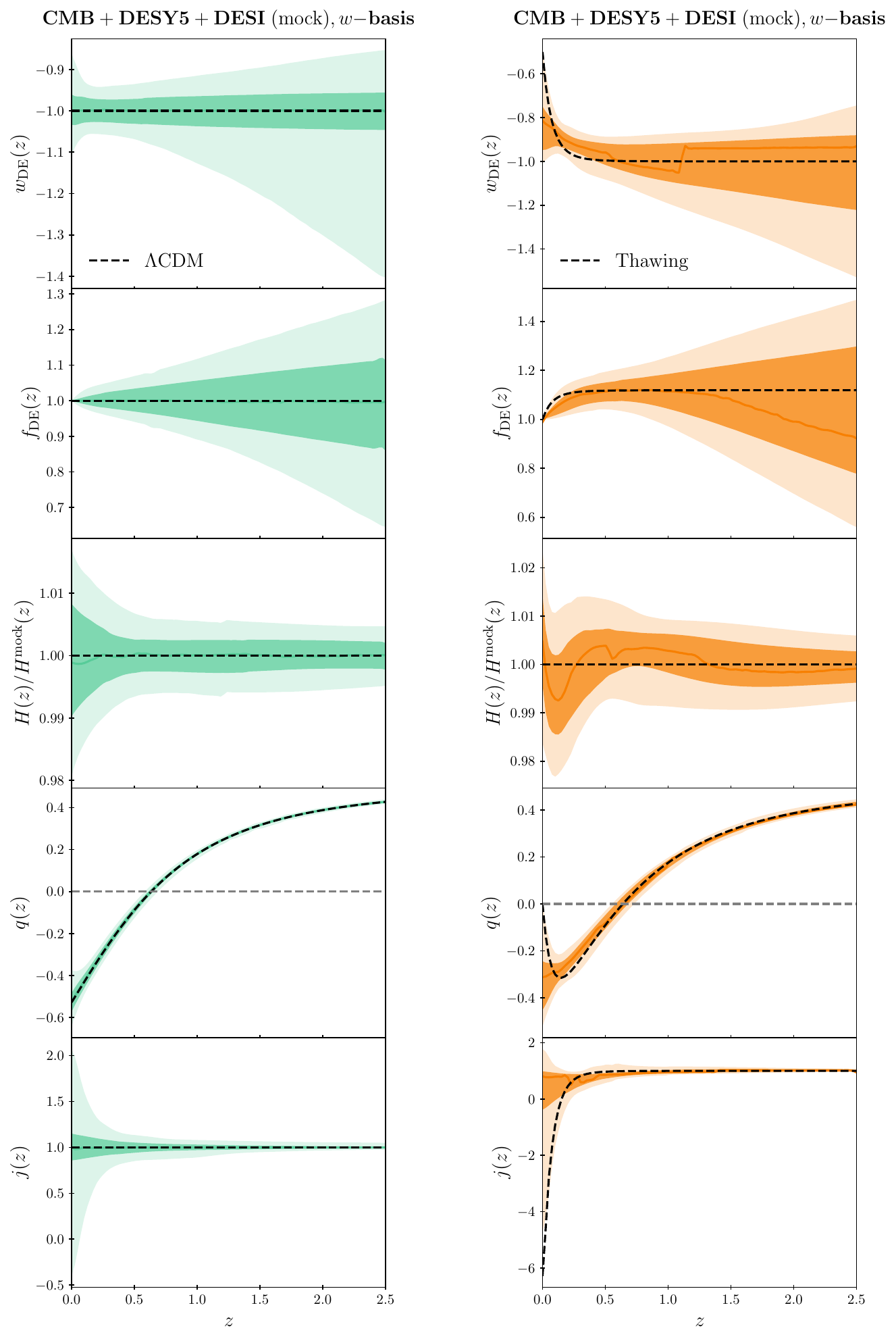}
    \caption{{\it Left plots:} Reconstructed functions obtained upon applying the WFR method with the $w$-basis to mock $\Lambda$CDM data. The black dashed curves correspond to the fiducial  $\Lambda$CDM model employed to build the aforementioned mock data set; {\it Right plots:} The same, but using the mock data generated assuming the thawing DE model of Eq. \eqref{eq:algebraicthwaing}. See Appendix \ref{sec:test} for further details.}
    \label{fig:mock}
\end{figure}

The results of our tests with mock data are displayed in figure \ref{fig:mock}. We are able to retrieve the shape of the underlying $\Lambda$CDM model employed to generate the mock data in the analysis of the left plots, and make also an excellent job in the reconstruction of the thawing quintessence model assumed in the analysis of the right plots. The latter has been deliberately taken to exhibit a very abrupt transition of the EoS parameter at redshift $z\lesssim 0.5$, which is extremely difficult to recover with standard reconstruction techniques given the current precision of the data and the scarcity of data points in that redshift range.  Our reconstructions for the thawing quintessence case above $z=0.1$ successfully encompass the true behavior of the cosmological functions at $<1\sigma$ CL. Below that redshift range, our reconstructed functions agree with the true ones at $<2\sigma$ CL, and even at $<1\sigma$ CL in the case of the Hubble function. This performance is far better than other reconstruction techniques explored in the literature. For instance, Gaussian Processes were employed in Appendix E of \cite{DESI:DR2extended} to reconstruct exactly the same model with a much less limited success, finding differences in the EoS parameter at $z\sim 0$ at the level of $\sim 5\sigma$.

These tests on mock data demonstrate the strong performance of the WFR method even in extreme cases, such as the one studied in this appendix. The reconstruction of dynamical DE models with a smoother evolution of the EoS parameter is achieved without any difficulty.

\section{Contour plots}\label{sec:contourplots}

In this appendix, we provide a compendium of contour plots with the results of the individual fitting analyses, obtained for every truncated expression (cf. section \ref{sec:RecoFunc}) and for every data set (cf. section \ref{sec:data}). They correspond to figures \ref{fig:contour_P18}, \ref{fig:contour_rho_panplus1} and \ref{fig:contour_rho_panplus2}, see below.

\begin{figure}[t!]
    \centering
    \includegraphics[width=\linewidth]{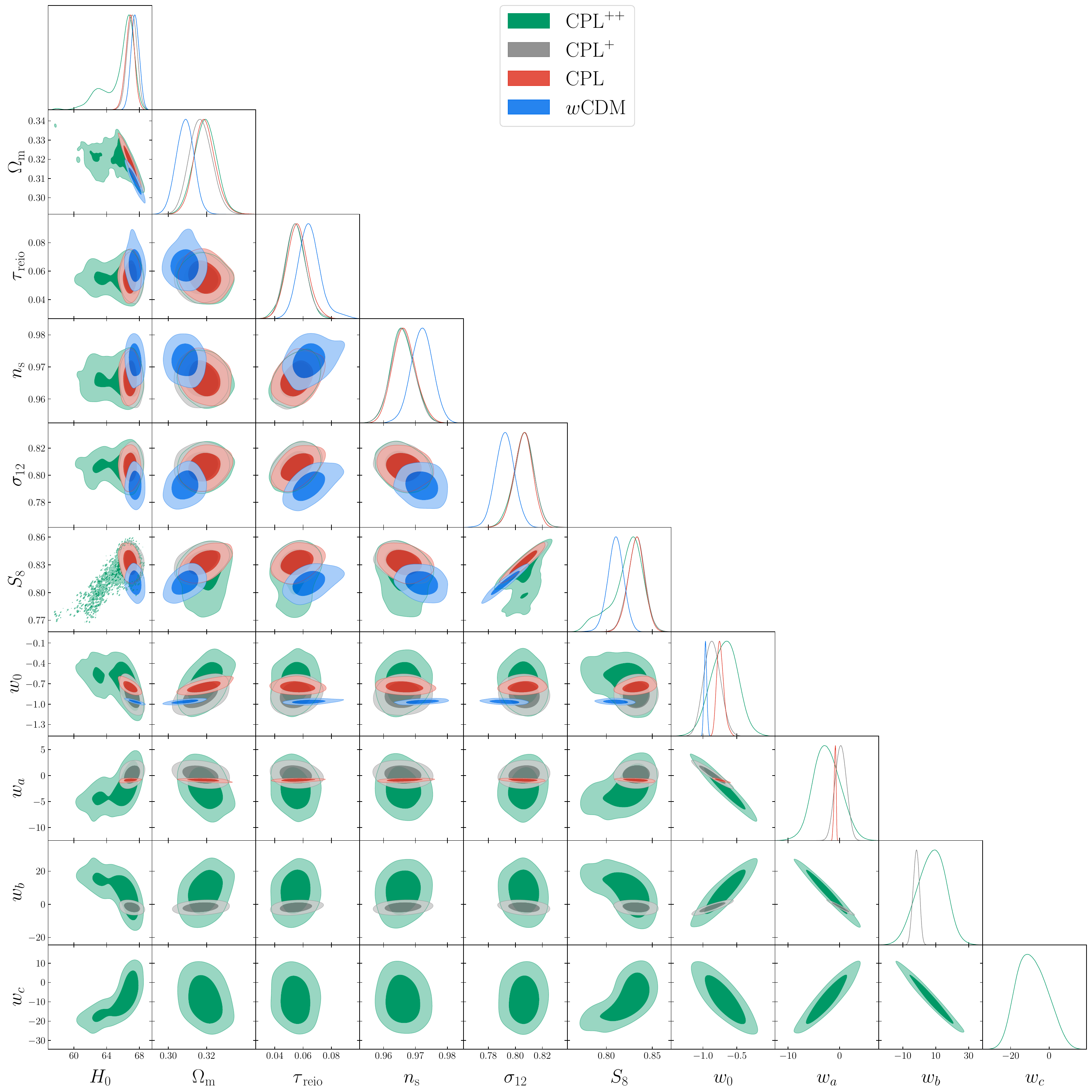}
    \caption{Contour plots at $68\%$ and $95\%$ CL for the combination Planck18+DES-Y5+DESI and using the $w$-basis. $H_0$ is given in km/s/Mpc.}
    \label{fig:contour_P18}
\end{figure}

\begin{figure}[t!]
    \centering
    \includegraphics[width=\linewidth]{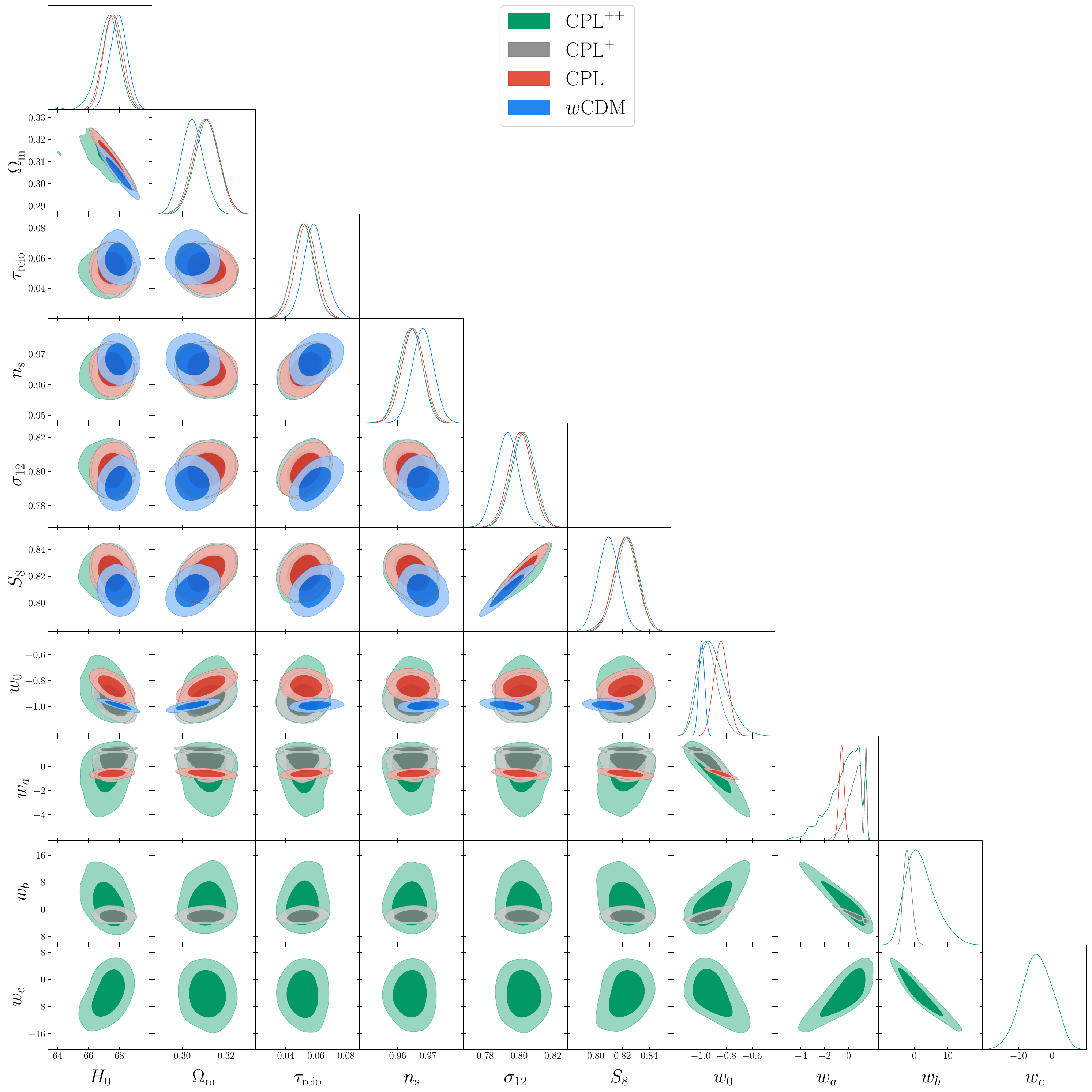}
    \caption{Contour plots at $68\%$ and $95\%$ CL for the combination PlanckPR4+PantheonPlus+DESI and using the $w$-basis. $H_0$ is given in km/s/Mpc.}
    \label{fig:contour_rho_panplus1}
\end{figure}

\begin{figure}[t!]
    \centering
    \includegraphics[width=\linewidth]{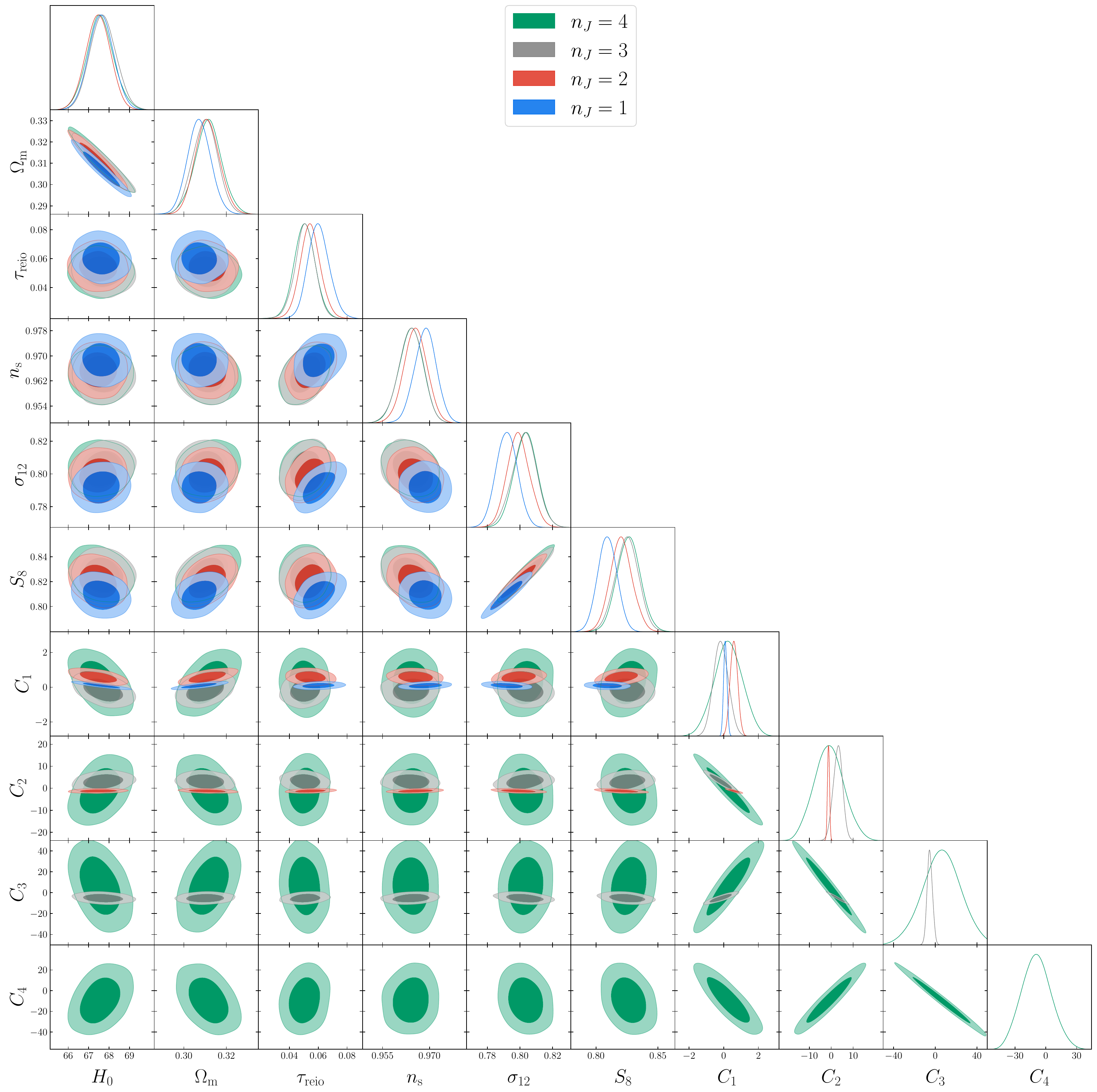}
    \caption{Contour plots at $68\%$ and $95\%$ CL for the combination PlanckPR4+PantheonPlus+DESI and using the $\rho$-basis. $H_0$ is given in km/s/Mpc.}
    \label{fig:contour_rho_panplus2}
\end{figure}


\section{AIC versus DIC for the WFR-weights}\label{sec:DIC}

In this appendix, we discuss the robustness of our results under the replacement of the information criteria employed to compute the weights in the WFR method. More concretely, we use the Deviance Information Criterion (DIC) \cite{DIC} to reconstruct the functions of interest and compare the results with those analyzed in the main body of the paper, which are obtained using the AIC. Let us first recall the reader the definition of DIC, 

\begin{equation}\label{eq:DIC}
   {\rm DIC} =  \chi^2(\bar{\theta})+2p_D\,,
\end{equation}
with $p_D=\overline{\chi^2}-\chi^2(\bar{\theta})$ the effective number of parameters in the model, $\overline{\chi^2}$ the mean value of $\chi^2$ and $\bar{\theta}$ the mean of the parameters entering the Monte Carlo analysis. It is important to stress that it incorporates the information encapsulated in the full Monte Carlo Markov chains.

\begin{figure}[t!]
    \centering
    \includegraphics[width=0.8\linewidth]{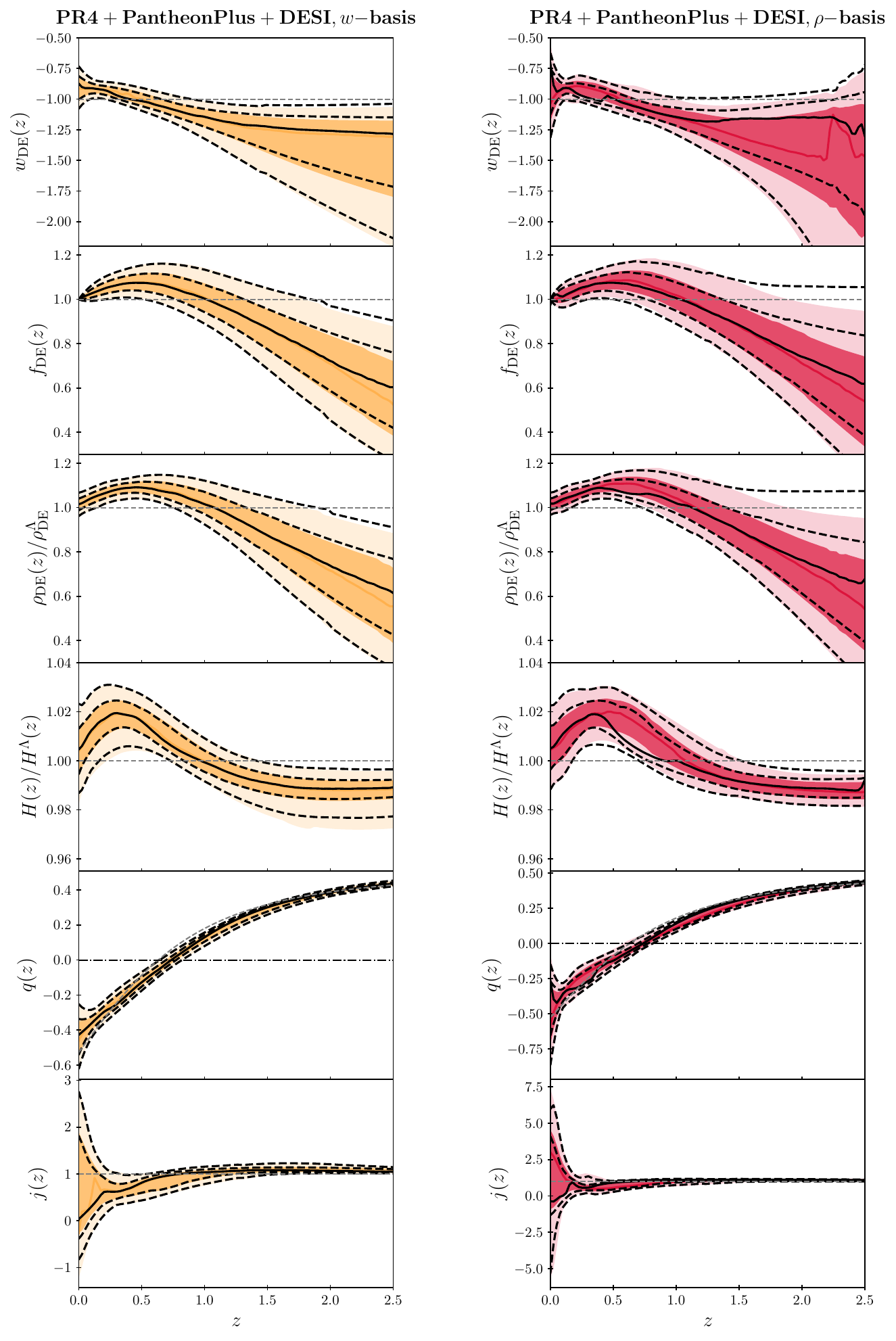}
    \caption{Comparison of reconstructed functions from DIC (colored) and AIC (black, cf. figure \ref{fig:reco}) with PR4+PantheonPlus+DESI using the $w$-basis (left column) and the $\rho$-basis (right column).}
    \label{fig:reco_DIC}
\end{figure}

Our results are displayed in Table \ref{tab:tabDIC} and in figure \ref{fig:reco_DIC}. They remain clearly stable. The constraints on the reconstructed parameters with DIC differ in almost all cases less than $0.2\sigma$ from those obtained with AIC, and the reconstructions of the various background functions are also in very good agreement. In the $\rho$-basis, the possibility of not having a crossing of the phantom divide is excluded from $2.08\sigma$ (with AIC) to $2.69\sigma$ (with DIC); in both cases the evidence remain quite below the $3\sigma$ level, while the probability of having negative dark energy density in the range $z\in[0,3]$ decreases slightly from 2.3\% to 2.1\% with AIC and DIC, respectively. In the $w$-basis, the stability of the results is even more evident. The absence of phantom divide crossing is rejected at $2.52\sigma$ (with AIC) and 2.73$\sigma$ (with DIC). The excellent agreement under the change of the information criterion employed to determine the weights, regardless of the WFR-basis, demonstrates the robustness of our analysis and conclusions.

Another information criterion one could use, in principle, is the Bayesian Information Criterion (BIC) \cite{BIC}. For a model $M_J$, it is defined as follows,

\begin{equation}
 {\rm BIC}_J = \chi^2_{{\rm min},J}+(n_c+n_J)\ln(n_{d})\,, 
\end{equation}
with $n_d$ the number of data points employed in the fitting analysis. Taking into account that we are considering $\mathcal{O}(10^3-10^4)$ data points, the complexity penalization factor reads in this case, $\ln(n_d)\simeq7-9$. This number is extremely large and leads to an unfair weighting of the various functions in the WFR-basis. For example, in order to contribute equally to the reconstruction under BIC, a model with one fewer parameter could yield a minimum $\chi^2$
 that is larger by 7–9 units compared to a model with an additional parameter. This is an undesirable feature, indicating that BIC is not suitable for the purposes of this study. In contrast, AIC and DIC provide a more balanced penalization of model complexity and yield fully compatible results.

\begin{table*}
\begin{center}
\begin{tabular}{|c|cc|}
\multicolumn{3}{c}{\textbf{Reconstructed parameters (DIC) }}\\ \hline
Parameter &  \textbf{$w$-basis} & \textbf{$\rho$-basis} \\
\hline
$10^2\omega_\mathrm{b}$ & $2.221\pm 0.013$ ($0.05\sigma$)  & $2.222\pm 0.014$ ($0.00\sigma$) \\ 
$10\,\omega_\mathrm{cdm}$ &  $1.191^{+0.009}_{-0.008}$ ($0.00\sigma$) & $1.192\pm 0.010$ ($0.07\sigma$) \\ 
$\ln(10^{10}A_s$) & $3.035\pm 0.014$ ($0.00\sigma$) & $3.033 \pm 0.014$ $(0.10\sigma)$ \\ 
$\tau$ & $0.052\pm 0.007$ ($0.00\sigma$) & $0.051\pm 0.007$ ($0.10\sigma$) \\ 
$n_s$ & $0.965\pm 0.004$ ($0.00\sigma$)  & $0.964\pm 0.004$ ($0.18\sigma$) \\ 
$H_0$ [kms/s/Mpc] & $67.48_{-0.63}^{+0.65}$($0.04\sigma$)  & $67.63 \pm 0.63$ ($0.02\sigma$) \\  \hline
$\Omega_\mathrm{m}^0$ &  $0.311\pm 0.006$ ($0.00\sigma$) & $0.311\pm 0.006$ ($0.00\sigma$)\\ 
$M$ [mag] & $-19.421\pm 0.015$ ($0.00\sigma$) & $-19.424\pm 0.015$ ($0.05\sigma$)\\ 
$\sigma_{12}$ & $0.802\pm 0.007$ ($0.10\sigma$) & $0.802^{+0.008}_{-0.007}$ ($0.09\sigma$)\\ 
$S_8$ & $0.823\pm 0.009$ ($0.00\sigma$) & $0.825\pm 0.010$ ($0.14\sigma$) \\
$z_{\rm t}$ & $0.74_{-0.05}^{+0.04}$ ($0.00\sigma$)  & $0.74 \pm 0.05$ ($0.00\sigma$)\\
$F_{\rm cross}$ [\%] & $99.37$ & $99.28$  \\
$F_{\rm neg}$ [\%] &  $-$ & $2.14$ \\
$z_{\rm cross}$ & $0.52_{-0.20}^{+0.12}$ ($0.66\sigma$) & $0.57^{+0.13}_{-0.20}$ ($0.00\sigma$) \\
$\Delta \mathcal{N}_{\rm eff}$ & $3.10$ & $3.21$ \\\hline

\end{tabular}
\caption{Same as Table \ref{tab:tab4} but using the DIC criterion and PlanckPR4+PantheonPlus+DESI. In parenthesis, we also display the relative deviation between AIC and DIC constraints (in absolute value), over the number of $\sigma=\sqrt{\sigma_{\rm AIC}^2+\sigma_{\rm DIC}^2}$ of each parameter. We make use of the results in the second and third columns of Table \ref{tab:tab4}. When the errors are asymmetric, we take the mean of the upper and lower uncertainties.}\label{tab:tabDIC}
\end{center}
\end{table*}


\acknowledgments The authors are funded by “la Caixa” Foundation (ID 100010434) and the European Union's Horizon 2020 research and innovation programme under the Marie Sklodowska-Curie grant agreement No 847648, with fellowship code LCF/BQ/PI23/11970027. They acknowledge the participation in the COST Action CA21136 “Addressing observational tensions in cosmology with systematics and fundamental physics” (CosmoVerse), and thank Prof. Joan Solà Peracaula and Prof. Arman Shafieloo for valuable discussions. They are also grateful to the anonymous referee for his/her useful suggestions, which motivated appendices A and C of this work.  


\end{document}